\documentclass[nonacm,sigconf]{acmart}

\setcopyright{none}
\renewcommand\footnotetextcopyrightpermission[1]{}
\makeatletter
\def\@acmSubmissionID{}
\def\@acmSubmissionType{}
\def\@journalName{}
\def\@acmPrice{}
\def\@copyrightpermission{}
\def\footers@default{}%
\makeatother

\usepackage{makecell}
\usepackage{graphicx}
\usepackage{subcaption}
\usepackage{csquotes}
\usepackage{balance}

\usepackage{xcolor}
\definecolor{darkbluebg}{HTML}{0F172A}
\definecolor{lightbluebg}{HTML}{1E293B}

\usepackage{tcolorbox}
\usepackage{makecell}
\usepackage{multirow}
\usepackage{pifont}
\AtBeginDocument{%
  }

\begin{document}

\title[Emergent Coordinated Behaviors in Networked LLM Agents]{Emergent Coordinated Behaviors in Networked LLM Agents: Modeling the Strategic Dynamics of Information Operations}


\author{Gian Marco Orlando}
\authornote{Both authors contributed equally to this research.}
\affiliation{%
  \institution{University of Naples Federico II}
  \department{Department of Electrical Engineering and Information Technology}
  \city{Napoli}
  \country{Italy}}
\email{gianmarco.orlando@unina.it}

\author{Jinyi Ye}
\authornotemark[1]
\affiliation{%
  \institution{University of Southern California}
  \department{Thomas Lord Department of Computer Science}
  \city{Los Angeles}
  \state{CA}
  \country{USA}}
\email{jinyiy@usc.edu}

\author{Valerio La Gatta}
\affiliation{%
  \institution{Northwestern University}
  \department{Department of Computer Science}
  \city{Evanston}
  \state{IL}
  \country{USA}}
\email{valerio.lagatta@northwestern.edu}

\author{Mahdi Saeedi}
\affiliation{%
  \institution{University of Southern California}
  \department{Information Sciences Institute}
  \city{Marina Del Rey}
  \state{CA}
  \country{USA}}
\email{mahdisae@usc.edu}

\author{Vincenzo Moscato}
\affiliation{%
  \institution{University of Naples Federico II}
  \department{Department of Electrical Engineering and Information Technology}
  \city{Napoli}
  \country{Italy}}
\email{vincenzo.moscato@unina.it}

\author{Emilio Ferrara}
\affiliation{%
  \institution{University of Southern California}
  \department{Thomas Lord Department of Computer Science}
  \city{Los Angeles}
  \state{CA}
  \country{USA}}
\email{emiliofe@usc.edu}

\author{Luca Luceri}
\affiliation{%
  \institution{University of Southern California}
  \department{Information Sciences Institute}
  \city{Marina Del Rey}
  \state{CA}
  \country{USA}}
\email{lluceri@isi.edu}

\renewcommand{\shortauthors}{Orlando et al.}

\begin{abstract}
Generative agents are rapidly advancing in sophistication, raising urgent questions about how they might coordinate when deployed in online ecosystems. This is particularly consequential in information operations (IOs), influence campaigns that aim to manipulate public opinion on social media. While traditional IOs have been orchestrated by human operators and relied on manually crafted tactics, agentic AI promises to make campaigns more automated, adaptive, and difficult to detect. This work presents the first systematic study of emergent coordination among generative agents in simulated IO campaigns. Using generative agent-based modeling, we instantiate IO and organic agents in a simulated environment and evaluate coordination across operational regimes, from simple goal alignment to team knowledge and collective decision-making. As operational regimes become more structured, IO networks become denser and more clustered, interactions more reciprocal and positive, narratives more homogeneous, amplification more synchronized, and hashtag adoption faster and more sustained.
% Notably, we find that merely informing agents about the identity of other agents sharing the same goals can generate coordination nearly as strong as when agents explicitly deliberate and vote on strategies.
Remarkably, simply revealing to agents which other agents share their goals can produce coordination levels nearly equivalent to those achieved through explicit deliberation and collective voting.
Overall, we show that generative agents, even without human guidance, can reproduce coordination strategies characteristic of real-world IOs, underscoring the societal risks posed by increasingly automated, self-organizing IOs.

% Emergent coordination among autonomous AI agents poses growing challenges for online ecosystems, especially in the context of information operations (IOs), orchestrated campaigns designed to manipulate public opinion. This study investigates how coordination can spontaneously emerge among Large Language Model (LLM)–driven agents engaged in a simulated IO campaign. Using generative multi-agent simulations as an experimental testbed, we systematically vary agents’ level of operational awareness (from shared goals to explicit Teammate Awareness and collective deliberation on strategies to guide subsequent actions) to study how structured awareness shapes collaborative behavior and influence diffusion.

% Our results show that increased awareness among IO agents markedly enhances internal organization, leading to denser and more reciprocal interaction networks, more homogeneous and emotionally aligned narratives, and more synchronized amplification behaviors through coordinated re-sharing tactics and campaign hashtag adoption. Notably, even simple mutual awareness among IO agents achieves a high degree of coordination, approaching the effectiveness of collective deliberation without requiring centralized planning.

% This work demonstrates that LLM-based generative agents can autonomously reproduce key coordination mechanisms observed in real-world IOs, underscoring both the risks posed by increasingly automated, self-organizing influence campaigns and the potential of generative multi-agent simulations as a methodological framework for studying emergent collective behavior.
\end{abstract}

%    <concept>
%        <concept_id>10003120.10003130.10003131.10011761</concept_id>
%        <concept_desc>Human-centered computing~Social media</concept_desc>
%        <concept_significance>500</concept_significance>
%        </concept>
%    <concept>
%        <concept_id>10010147.10010178.10010219.10010220</concept_id>
%        <concept_desc>Computing methodologies~Multi-agent systems</concept_desc>
%        <concept_significance>500</concept_significance>
%        </concept>
%  </ccs2012>

% \ccsdesc[500]{Human-centered computing~Social media}
% \ccsdesc[500]{Computing methodologies~Multi-agent systems}



\maketitle

\section{Introduction}

AI agents, often referred to as \textit{generative agents}, are autonomous entities capable of reasoning and interacting with minimal human supervision \cite{park2023generative}. As their sophistication increases, so does the potential for their large-scale deployment across online ecosystems. While this opens up new opportunities, it also raises critical questions about how such agents might coordinate when pursuing shared objectives including, potentially, nefarious ones.

A particularly high-stakes domain where coordination dynamics play a central role is online information operations (IOs). IOs are orchestrated influence campaigns that seek to manipulate public opinion on social media, often in the context of geopolitical issues or societally-relevant events (elections, crises, etc.) \cite{badawy2018analyzing, cinus2025, minici2024uncovering}. Traditional IOs have typically been orchestrated by human operators, relying on manually crafted coordination strategies executed through both human- and software-controlled accounts \cite{ferrara2016rise}. Despite this hybrid organization, real-world campaigns have generally employed relatively simple tactics, such as synchronized posting, hashtag flooding, or retweet rings, to create the illusion of widespread consensus and manipulate content recommendation algorithms \cite{Starbird2019Disinformation, Arif2018Acting, Badawy2019Anatomy}.

With the advent of large language models (LLMs) and agentic AI, IOs are expected to grow far more sophisticated: campaigns may become largely automated, highly adaptive, and capable of self-organized coordination spanning large networks of AI agents with minimal or no human oversight. Understanding the mechanisms and consequences of emergent, automated coordination among generative agents is therefore a pressing research problem, with direct implications for information integrity and platform governance. 
% This raises an urgent research question: \textit{If LLM-powered agents are deployed in IO campaigns, how might coordination emerge, and to what extent do these strategies mirror those observed in real-world IOs?}
This raises an urgent research question: \textit{If generative agents are employed in information operations, how does coordination arise among them, and to what extent do their strategies resemble those observed in real-world campaigns?}

\paragraph{\textbf{Contribution of this work.}}
To address this question, we simulate an IO campaign in which generative agents act as organic users and IO operators within a shared online environment. In this simulated campaign, IO agents seek to promote a political candidate and amplify a shared hashtag across the network. We employ Generative Agent-Based Modeling (GABM) \cite{ghaffarzadegan2024generative} as our methodological framework to simulate multi-agent interactions and examine whether coordination patterns naturally emerge. 

To investigate this, we examine three progressively structured \textit{operational regimes}: (i) \textit{Common Goal}, where IO agents share only the high-level objective of the IO but lack awareness of their teammates and shared coordination strategies; (ii) \textit{Teammate Awareness}, where agents are explicitly informed of their allies’ identities and can potentially support each other in their common goal; and (iii) \textit{Collective Decision-Making}, where agents periodically deliberate and vote on strategies to guide subsequent actions. Building on empirical findings from real-world IO studies (e.g., \cite{Luceri2024Deceit, Pacheco2021Coordinated, Arif2018Acting}), this experimental design is guided by a set of hypotheses that map observed coordination mechanisms into measurable behavioral dimensions within the simulation. In particular, we propose and evaluate both \emph{coordination metrics}, which quantify the scale, intensity, and temporal dynamics of collaboration among IO agents, and \emph{impact metrics}, which capture the engagement garnered by IO agents from organic agents and the diffusion of their promoted hashtag across the network. Importantly, our analysis prioritizes the mechanisms underlying coordination, rather than the political dimensions of the simulated campaign, emphasizing how increasing levels of operational awareness drive the emergence and amplification of coordination dynamics among generative agents with no human intervention.
Notably, we find that minimal information about peer agents can trigger coordination nearly as strong as when agents collaboratively deliberate on strategies.
% Our contributions are threefold. First, we introduce an empirically grounded hypothesis framework, derived from real-world IO studies, to assess coordination among LLM-driven agents. Second, applying this framework across increasingly structured operational regimes---from shared goals to explicit Teammate Awareness and collective deliberation---we provide a systematic study of emergent coordinated behaviors, showing that agents transition from loosely aligned to tightly organized activity. 
Specifically, IO networks become denser and more clustered; interactions grow more reciprocal and increasingly positive; narratives converge toward greater semantic homogeneity, with amplification occurring through more synchronized re-sharing; and campaign-hashtag adoption accelerates earlier and sustains higher cumulative uptake. 
% revealing only who shares their goal allows agents to coordinate almost as strongly as if they had engaged in explicit deliberation.

Finally, we release an interactive dashboard\footnote{\url{https://llmxio-dashboard.vercel.app/}} that enables real-time exploration of coordination dynamics as they unfold within the simulated environment. The dashboard visualizes the evolving social graph, including comment and re-share networks, alongside the content agents produce and engage with. Importantly, it provides researchers with a powerful resource to examine coordination dynamics in simulated IOs, highlighting key indicators, such as the trajectory of hashtag adoption among organic agents, and uncovering the motivations driving coordinated behavior. In this paper, we provide qualitative insights into IO agents’ strategic decision-making and their articulated motivations, while additional details on the dashboard’s functionality are presented in the \textit{Appendix}. %By making these dynamics transparent and explorable, the dashboard serves as both an analytical tool for validating hypotheses and a reproducible artifact that enables the broader community to probe multi-agent coordination mechanisms in generative simulations.

Overall, this work provides the first demonstration that generative agents can autonomously reproduce key coordination mechanisms characteristic of real-world IO campaigns, operating without human intervention. In doing so, it highlights the societal risks posed by increasingly automated IOs on social systems. To ensure transparency and reproducibility, we publicly release our code\footnote{\url{https://anonymous.4open.science/r/GABMxIO-D546/}}.
% and the potential of GABM as a methodology for studying emergent multi-agent coordination.

\section{Related Work}

\subsection{Emerging Collective Behaviors in LLM-based Social Simulations}

Recent advances in LLM-based multi-agent simulations demonstrate that generative agents can exhibit emergent collective behaviors in cooperative, competitive, and communicative contexts. Prior studies show that language-enabled agents can negotiate, form conventions, and coordinate around predefined tasks like games and decision-making \cite{ashery2025emergent, chen2025multiagentconsensusseekinglarge, vallinder2024culturalevolutioncooperationllm, buscemi2025strategiccommunicationlanguagebias, tran2025multi}. These results collectively suggest that LLM agents can potentially develop shared conventions and joint strategies without explicit rules. 

However, one open question is how coordination emerges and develops within a group of autonomous agents with a common objective under open-ended, unstructured conditions. Within this context, coordination through social influence, where agents align narratives or behaviors to shape others' beliefs or actions, remain insufficiently explored. 

While several works simulate opinion dynamics and polarization \cite{chuang-etal-2024-simulating, piao2025emergencehumanlikepolarizationlarge,cau2025languagedrivenopiniondynamicsagentbased}, or adversarial collusion such as misinformation diffusion and financial fraud \cite{Ren2025Collusion, Qiao2025BotSim, ng2025llm}, two limitations remain. First, coordination among agents is typically predefined rather than emergent. Second, evaluation often relies on narrow outcome metrics (e.g., engagement counts, sentiment shifts). Therefore, they lack tracing the dynamic processes through which coordination structures form and evolve. Furthermore, recent studies on model-level alignment and information suppression highlight how internal moderation constraints can shape what information agents choose to share or omit \cite{qiu2026information}, which in turn may influence collective behaviors simulated in multi-agent environments.

Our study addresses these gaps by simulating an information operation with GABM and systematically varying levels of operational awareness, allowing us to investigate \emph{how coordination strategies arise naturally}. By coupling behavioral, network, and diffusion-based metrics, we provide a richer, process-level understanding of how organized coordination behaviors arise and propagate within LLM-driven social systems.

\subsection{Empirical Studies of Online IOs}
Several studies document coordinated activity driving IOs on online platforms, revealing concrete strategies such as \emph{synchronized posting} and temporally clustered behaviors \cite{pacheco2020unveiling, Pacheco2021Coordinated, cinus2025}; \emph{hashtag flooding} and narrative amplification through co-occurring tags \cite{Arif2018Acting, Luceri2024Deceit, luceri2025coordinated, minici2024uncovering}; \emph{retweet} (or re-share) \emph{rings} that generate artificial popularity signals \cite{Pacheco2021Coordinated, Luceri_Co_Retweet}; and \emph{coordinated reply attacks} that target influential accounts to steer audience perception in the comment space \cite{Pote2025Reply}. This suite of tactics is commonly employed to create the illusion of public consensus around certain viewpoints and to game platform recommendation systems, making content appear more viral than it truly is. Additional evidence shows that these coordinated behaviors often arise from \emph{collaborative work} between human- and automated-controlled accounts following scripted strategic actions, rather than adaptive or deliberative strategy formation \cite{keller2020political, kumar2017army}. Guided by these observations, we next present our research hypotheses and methodology.
\section{Research Hypotheses \& Methodology}

To systematically evaluate coordination behaviors among networked, generative agents, we ground our work in empirical findings from previous studies on online IOs. We formulate testable hypotheses that map real-world coordination signals to measurable metrics within our generative agent-based simulation.
Our hypotheses are organized around two complementary dimensions: the strategic \textit{coordination} among IO agents (H1–H3) and their resulting \textit{impact} on organic agents (H4–H5). The coordination dimension captures how IO agents autonomously align and reinforce one another through network cohesion, narrative convergence, and re-share amplification, without human intervention. The impact dimension evaluates how these coordinated behaviors translate into influence outcomes, reflected in organic agents’ adoption of a promoted hashtag, their engagement patterns with IO agents, audience diversity, and the size, depth, and breadth of resulting information cascades. Together, these dimensions provide a structured framework for comparing simulated coordination dynamics against empirical patterns documented in real-world IO campaigns.

We posit that increasing levels of operational regimes,\footnote{We use the terms \textit{operational regimes} and \textit{operational awareness} interchangeably to denote settings in which IO agents have progressively greater knowledge of their goals, teammates, and strategies.} ranging from basic goal knowledge to team composition and strategy deliberation, will progressively strengthen coordination mechanisms and amplify campaign impact. To systematically validate this overarching thesis, we formalize five specific hypotheses that operationalize observable coordination patterns and impact metrics.
% : H1–H3 examine internal coordination mechanisms among IO agents, while H4–H5 assess the external influence these coordinated behaviors exert on organic user populations.

\paragraph{\textbf{H1: Network Cohesion.}}
Empirical IO campaigns often exhibit dense clusters of interactions among coordinated accounts \cite{Luceri2024, minici2024uncovering, cinus2025, vargas2020detection}. We therefore hypothesize that increasing operational awareness among IO agents will translate into denser networks and more reciprocal interactions among IO agents. 

\textit{Operationalization.}
To test H1, we analyze the evolution of follow, comment, and re-share networks between IO and organic agents, and quantify intra-group network properties within the IO community, including \textit{network density}, \textit{clustering coefficient}, and \textit{reciprocity}, which together capture how tightly IO agents coordinate and mutually reinforce one another within the social network.

\paragraph{\textbf{H2: Narrative Convergence.}}
Prior research shows that coordinated actors reinforce a shared narrative frame through repeated talking points, hashtags, or slogans \cite{Arif2018Acting, Pacheco2021Coordinated, pacheco2020unveiling}, thereby creating the impression of broad consensus. We therefore hypothesize that increasing levels of operational regimes will lead to stronger convergence not only in the narratives IO agents propagate but also in the sentiment with which these narratives are expressed \cite{burghardt2024socio}.

\textit{Operationalization.}
To test H2, we measure both \textit{textual} and \textit{affective convergence} within the IO community. 
Following methods used in empirical coordination studies \cite{ng2023coordinating, vishnuprasad2024tracking}, we compute pairwise cosine similarity between IO agents' posts, where text embeddings are obtained using Sentence-BERT.
%normalized against the similarities observed in the organic agent group. 
% We then track whether the mean intra-group similarity increases with higher levels of operational awareness. 
In parallel, we assess affective convergence using transformer-based sentiment classification on IO-to-IO comments, applying a RoBERTa-based sentiment classifier fine-tuned on social media text \cite{barbieri2020tweeteval}.
% \footnote{\url{https://huggingface.co/cardiffnlp/twitter-roberta-base-sentiment}}.
% Predicted sentiment scores are mapped to $\{-1,0,+1\}$ and averaged across runs to quantify the degree of positive affect expressed toward teammates. 
Increasing textual similarity and positive sentiment across operational regimes would indicate stronger narrative and affective alignment among IO agents.

\paragraph{\textbf{H3: Amplification through Re-sharing Behaviors.}}
Coordinated re-sharing is a common tactic in influence campaigns to artificially amplify content (e.g., tweets, hashtags, or URLs), making it appear more viral and credible than it would be organically \cite{Pacheco2021Coordinated, hristakieva2022, Luceri2024Deceit, pacheco2020unveiling, schoch2022coordination}. 
We hypothesize that increasing operational awareness among IO agents enhances the degree to which they systematically re-share similar content, thereby amplifying message visibility and reinforcing narrative dominance.

\textit{Operationalization.}
To test H3, we quantify coordination strength using a \textit{co-retweet similarity} measure, following approaches adopted in empirical IO studies \cite{Luceri_Co_Retweet}. 
For each simulation run, we build a bipartite graph linking IO agents to the original posts they re-shared and compute pairwise cosine similarity between agents' TF-IDF vectors to quantify overlap in amplified content. Higher values reflect tighter synchronization of re-sharing behaviors under increased operational awareness.

\paragraph{\noindent\textbf{H4: Hashtag Adoption.}}
A central tactic of IOs is the repeated posting and amplification of promoted hashtags to dominate online discourse and manipulate platform trending algorithms \cite{Arif2018Acting, Luceri2024Deceit, luceri2025coordinated, minici2024uncovering}. 
We hypothesize that increasing operational awareness among IO agents enhances the overall adoption and diffusion of promoted hashtags from IO agents to the wider organic agent base.

\textit{Operationalization.}
To test H4, we introduce a campaign-specific hashtag accessible only to IO agents at initialization. 
Throughout the simulation, IO agents are tasked with maximizing its visibility, while organic agents may adopt it through direct interaction or indirect exposure. 
We quantify diffusion outcomes using three complementary measures: 
(i) the proportion of posts containing the hashtag and the proportion of organic agents adopting the hashtag (via original posts or re-shares); 
(ii) the time lag between each organic agent's first interaction with an IO agent and their first hashtag adoption; and 
(iii) the number of exposures to the hashtag before first adoption, following the approach in \cite{ye2024susceptibility}.

\paragraph{\textbf{H5: Cross-Group Diffusion.}}
Beyond internal echoing, coordinated campaigns are expected to facilitate the diffusion of IO-generated content across community boundaries \cite{dey2024coordinated}. 
We hypothesize that higher levels of operational regimes will increase the extent and heterogeneity of organic engagement with IO agents, leading to broader audience reach and more extensive content propagation.

\textit{Operationalization.}
To evaluate H5, we measure three indicators of diffusion: (i) \textit{engagement counts}, defined as the number of retweets and comments that IO agents receive from organic agents; (ii) \textit{audience diversity}, which captures how heterogenous the IO agents' audience is. The Gini coefficient is a suitable metric for this purpose, as it quantifies the concentration and inequality of interactions across users \cite{van2016employing, ye2025auditing}.
For each IO agent, we count the number of interactions received from each unique organic agent and compute the Gini coefficient $G$ of these interaction counts.  
% The coefficient is given by
% \[
% G = \frac{\sum_{i=1}^{n} (2i - n - 1)c_i}{n \sum_{i=1}^{n} c_i},
% \]
% where $c_i$ denotes the number of interactions from organic agent $i$ and $n$ is the number of unique organic agents.  
Lower $G$ values indicate evenly distributed engagement, while higher values indicate concentration among a few agents.  
We report the diversity score as $1 - G$, such that higher values reflect broader and more heterogeneous audience reach; and (iii) \textit{cascade magnitude}, which captures the structural extent of IO-generated content diffusion.
For each IO-initiated tweet, we reconstruct its full diffusion cascade (via re-share and comment) and compute its \textit{size} (total number of tweets in the cascade), \textit{depth} (longest root-to-leaf path), and \textit{breadth} (maximum number of nodes appearing at the same cascade level), reflecting how far, how widely, and how extensively IO-generated contents spread.

\begin{figure*}[t]
    \centering
    \resizebox{0.9\textwidth}{!}{%
        \begin{tabular}{ccc}
            \begin{subfigure}[t]{0.32\textwidth}
                \centering
                \includegraphics[width=\linewidth]{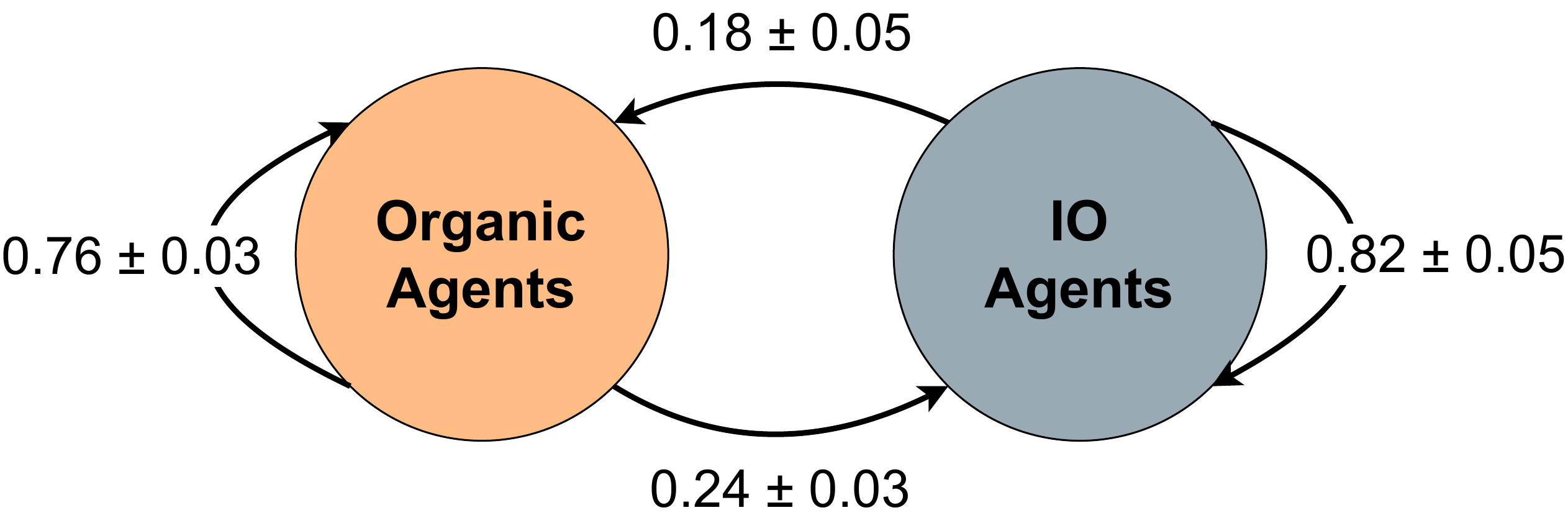}
                \caption{Common Goal}
                \label{fig:retweet_narrative_only_hashtag}
            \end{subfigure} &
            \begin{subfigure}[t]{0.32\textwidth}
                \centering
                \includegraphics[width=\linewidth]{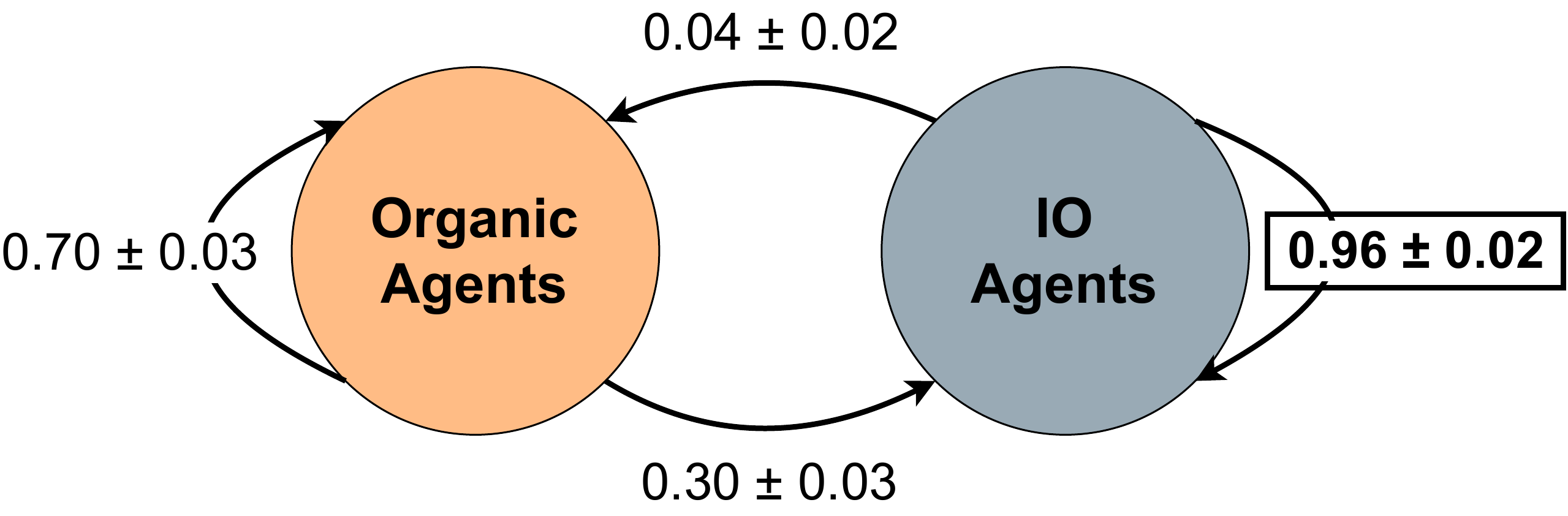}
                \caption{Teammate Awareness}
                \label{fig:retweet_knowing_mates_hashtag}
            \end{subfigure} &
            \begin{subfigure}[t]{0.32\textwidth}
                \centering
                \includegraphics[width=\linewidth]{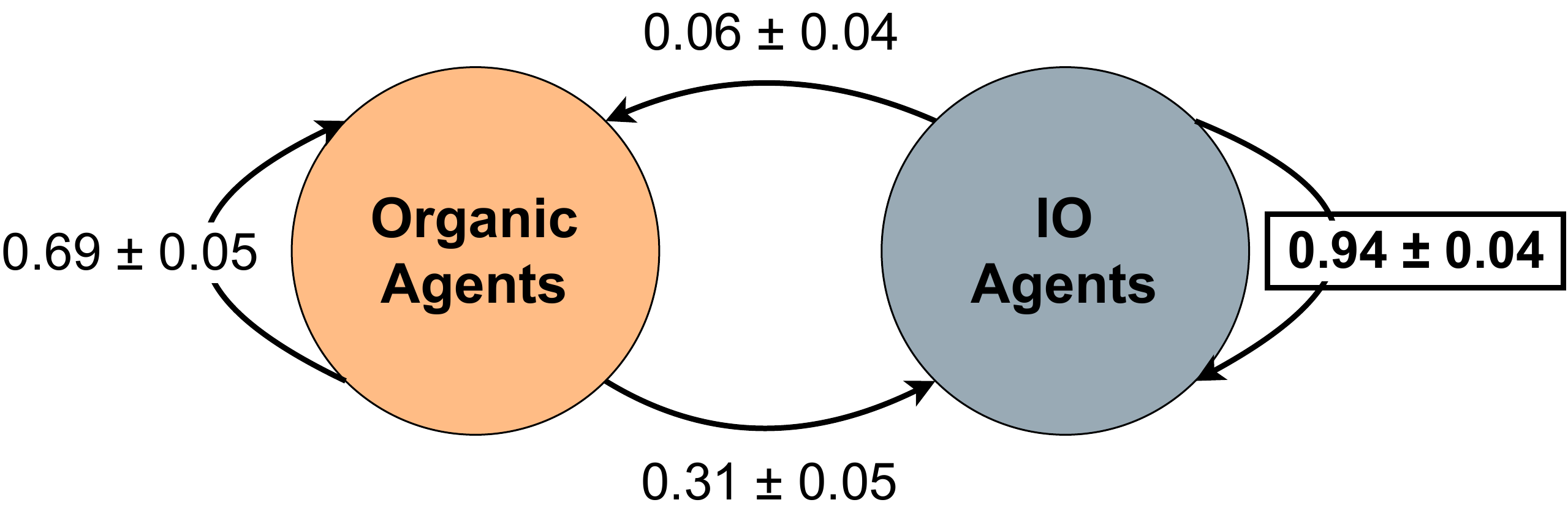}
                \caption{Collective Decision-Making}
                \label{fig:retweet_collective_decision_hashtag}
            \end{subfigure}
        \end{tabular}%
    }
    \caption{Re-share network across operational settings. Intra-group amplification among IO agents increases with operational awareness. Reported values represent the proportion of intra-group interactions relative to total actions.}
    \label{fig:retweet_networks_hashtag}
\end{figure*}

\section{Experimental Setup}
\label{sec:experimental_setup}

\subsection{Simulation Framework}

We employ the agent-based simulation framework from \cite{ferraro2024agent}, which models social media platforms like Twitter/X as dynamic ecosystems of heterogeneous LLM-powered agents. Each agent comprises three core components: a persona encoding its identity and group affiliation, a memory module storing interaction history, and an action policy that autonomously determines behaviors (posting, commenting, re-sharing, following) by integrating persona-based preferences with environmental feedback. The environment includes an evolving network topology shaped by following behaviors and a recommender system regulating content exposure. This framework has demonstrated capability in capturing emergent phenomena such as polarization and information diffusion \cite{ferraro2024agent,orlando2025can}.

The simulation proceeds iteratively over fixed timesteps. At each iteration, agents receive personalized content recommendations and autonomously select actions (e.g., original post or re-share) based on feedback from prior posts (e.g., received re-shares or comments), activity patterns from memory, and available recommended content. After all agents act, the environment updates, processes engagement metrics, and advances to the next iteration. For our experiments, each simulation runs for 50 iterations with three repetitions per configuration to account for stochastic variability. Implementation details regarding the recommender system, memory updates, and environment feedback are provided in \textit{Appendix}~\ref{app:implementation_details}.

Each simulation involves 50 agents: 10 IO actors and 40 organic agents. IO agents are instructed via system prompts to conduct a political influence campaign aimed at promoting a political candidate and maximizing the adoption of a campaign-specific hashtag, which is initially known only within the IO group. The information available to IO agents, including teammate knowledge and access to collective strategy discussions, varies across operational settings, as described in the next section. The 40 organic agents are evenly divided based on their political alignment with the IO campaign's messaging: 20 \textit{aligned} organic agents, whose viewpoints align with the promoted perspective, and 20 \textit{not aligned} organic agents, whose viewpoints oppose it. Following \cite{ferraro2024agent}, these agents are instantiated with preferences and affiliations derived from real Twitter users. Specifically, organic agent profiles are initialized using the U.S. 2020 Election dataset \cite{US_Dataset}, leveraging the annotations from \cite{ferraro2024agent} to distinguish the two classes of users with opposing political perspectives. Full prompt details are provided in \textit{Appendix}~\ref{app:implementation_details}. 

% \paragraph{Content recommendation system.}

% Content exposure was driven by a recommendation system designed to emulate key structural and algorithmic properties of real-world platforms. At each iteration, agents received up to 100 content recommendations, 60\% of these drawn from in-network sources (i.e., accounts that the agent followed) and the remaining 40\% randomly sampled from out-of-network agents. 

% \paragraph{Activation probabilities.}
% % Maybe appendix
% Actions—including following, posting, re-sharing, liking, and commenting—were determined through probabilistic sampling. A random number generator with a threshold of 0.5 governed action selection, while the underlying LLM generated the content associated with each action.

% \paragraph{Implementation details.}
% % Maybe appendix
% The experimental framework was implemented using the PyAutogen library \cite{wu2024autogen}, with generative agents powered by the Llama~3.3~70B model\footnote{\url{https://ollama.com/library/llama3.3:70b}}. 
% Simulations were executed on a computing system equipped with 2$\times$A100/A40/V100 GPUs. 
% Each \textit{simulation iteration} required approximately 60 minutes to complete under our configuration.
% To ensure transparency and reproducibility, we publicly release our code\footnote{\url{https://anonymous.4open.science/r/GABMxIO-D546/}}.

\subsection{Operational Regimes}

Beyond their core initialization as IO actors, we define three progressively structured regimes that vary the level of operational awareness among IO agents. It is worth noting that in none of these settings are agents guided by humans in selecting their actions, nor are they provided with explicit coordination guidelines. The regimes described below complement the agents' base instructions solely by modulating the information available to them.

\textbf{Common Goal}: In this baseline setting, IO agents are instructed via a prompt about their shared objective, i.e., to promote a political candidate and amplify the campaign hashtag. Each agent seeks to advance this objective without direct awareness of other participants, as they lack explicit knowledge of their teammates' identities. Coordination, if it emerges, arises implicitly through aligned goals rather than through deliberate collaboration.

\textbf{Teammate Awareness}: In this setting, IO agents are instructed about their shared objective and explicitly informed of the identities of their IO partners via system prompts. 
While each agent retains individual autonomy in tactical decision-making, this awareness may enable more targeted amplification strategies and direct mutual support (e.g., strategically re-sharing teammates' content). 
% Coordination may emerge as intentional rather than coincidental.

\textbf{Collective Decision-Making}: This setting introduces the most sophisticated operational regime. Every five time steps, all IO agents enter a private discussion channel where they are presented with detailed performance materials from the previous window, including individual and aggregated summaries of recent posts, engagement metrics, and recent IO-IO interactions. Inspired by the \textit{Reflection Module} in~\cite{park2023generative}, where agents periodically review experiences and synthesize insights to guide future behavior, this reflective step allows agents to evaluate collective outcomes and adapt coordination strategies based on shared situational feedback. Each agent independently proposes three recommendations for the next period, which are collected and consolidated by an \textit{IO Orchestrator}, an independent agent that identifies recurring themes, quantifies their frequency across IO agents, and ranks the top five actionable strategies to adopt based on these recommendations. The resulting collective strategy is shared back with all IO agents, who operationalize, refine, and update it in subsequent discussion cycles as new performance signals and coordination patterns emerge.

% To examine how coordination strength shapes campaign dynamics, we define three progressively stronger settings of operational awareness for IO agents. In the \textit{Common Goal} setting, agents are instructed to promote a shared narrative but lack explicit awareness of their teammates. In the \textit{Teammate Awareness} setting, agents are explicitly informed about their coordination partners, enabling more targeted amplification and direct interaction while preserving individual autonomy on strategic decision-making. Finally, in the \textit{Collective Decision-Making} setting, IO agents periodically enter a private discussion channel outside the simulated social media environment to deliberate, propose, and vote on campaign strategies (e.g., content creation, re-sharing priorities, engagement tactics) before returning to the public space. This design provides a structured progression from implicit alignment to explicit collaboration and collective deliberation, allowing us to systematically investigate how operational awareness influence both internal cohesion and campaign impact.

\section{Results}
This section presents the empirical results from our simulations. While the findings are structured according to their respective hypotheses, they are derived from a unified set of simulations rather than independent runs. Each operational setting was executed with \textit{three} independent repetitions to ensure robustness and account for stochastic variability.

\subsection{Network Cohesion (H1)} \label{sec:H1}

We test H1 by examining how different operational settings shape cohesion in the re-share, comment, and follow networks. A consistent pattern emerges: when IO agents are informed of their teammates, intra-group coordination increases. In the \textbf{re-share network}, the average proportion of re-shares targeting IO peers significantly increases from 0.82 in the \textit{Common Goal} setting to 0.96 in the \textit{Teammate Awareness} setting and remains high at 0.94 in the \textit{Collective Decision-Making} setting (Figures~\ref{fig:retweet_narrative_only_hashtag}--\ref{fig:retweet_collective_decision_hashtag}). The \textbf{comment network} shows a similar increase in intra-group exchanges, from 0.48 to 0.63 and 0.65 across the three regimes (Figure ~\ref{fig:comment_networks_hashtag} in \textit{Appendix}). In the \textbf{follow network}, the average proportion of within-group ties also grows significantly from 0.27 to 0.35 (Figure ~\ref{fig:follow_networks_hashtag} in \textit{Appendix}). 
% while cross-group ties remain largely stable
% It is worth noting that these values are computed by normalizing the intra-group interactions by the total number of actions of that type. 
Note that we do not highlight statistical differences among the metrics in this hypothesis, as the small sample size (three data points per setting) reduces statistical power and limits the reliability of significance testing. However, the small standard deviations indicate low variability across runs, suggesting stable and consistent coordination patterns within each setting.
The higher values observed in the \textit{Teammate Awareness} and \textit{Collective Decision-Making} settings suggest that, under more structured operational regimes, IO agents are more likely to follow, interact, and amplify one another, resulting in more cohesive IO networks.

% in the re-share network is marginal, the absolute level of intra-group re-sharing remains remarkably high when coordination conditions rise, highlighting that IO agents consistently prioritize amplifying one another’s content.

The increased cohesion of the IO community is further captured by intra-group connectivity metrics derived from a directed network aggregating all comment and re-share interactions among IO agents. We compute density, clustering coefficient, and reciprocity to characterize internal coordination structures. As shown in Table~\ref{tab:coordination_metrics}, more structured operational regimes yield consistently stronger intra-group cohesion. Mean network density increases from 0.74 in the \textit{Common Goal} condition to 0.89 in both \textit{Teammate Awareness} and \textit{Collective Decision-Making} settings, while the clustering coefficient rises from 0.86 to 0.96 and 0.97. Reciprocity also improves, from 0.56 to 0.68, indicating that ties become more mutual when agents are explicitly aware of their peers.

\textbf{Summary (H1).} Together, these results support H1, indicating that higher levels of operational awareness foster denser, more clustered, and more reciprocal IO networks.

\begin{table}[t]
    \centering
    \caption{\textit{Coordination metrics} of the IO intra-group network under different operational settings. Reported values represent means across three simulation runs, with the standard deviation of the mean shown in parentheses. The highest value among the three settings is bolded, and the second-highest is underlined. Asterisks (***) denote statistically significant differences from the \textit{Common Goal} condition (Mann–Whitney U test, $^{***}\!:\,p<0.001$). For content alignment and amplification metrics, results for IO agents are shown alongside the corresponding organic agent baseline.}
    \label{tab:coordination_metrics}
    \resizebox{\columnwidth}{!}{%
    \begin{tabular}{lccc}
        \toprule
        & \makecell{\textbf{Common} \\ \textbf{Goal}} 
        & \makecell{\textbf{Teammate} \\ \textbf{Awareness}} 
        & \makecell{\textbf{Collective} \\ \textbf{Decision-Making}} \\
        \midrule
        \multicolumn{4}{l}{\textit{Intra-Cluster Metrics (IO Agents)}} \\
        \midrule
        Density (\textbf{H1})
        & \makecell{0.74 \\ (± 0.05)} 
        & \makecell{\textbf{0.89} \\ (± 0.03)} 
        & \makecell{\textbf{0.89} \\ (± 0.02)} \\
        Clustering Coefficient (\textbf{H1})
        & \makecell{0.86 \\ (± 0.01)} 
        & \makecell{\underline{0.96} \\ (± 0.02)} 
        & \makecell{\textbf{0.97} \\ (± 0.03)} \\
        Reciprocity (\textbf{H1})
        & \makecell{0.56 \\ (± 0.07)}
        & \makecell{\textbf{0.68} \\ (± 0.07)}
        & \makecell{\underline{0.65} \\ (± 0.07)} \\
        \midrule
        \multicolumn{4}{l}{\textit{Content Alignment and Amplification (IO Agents)}} \\
        \midrule
        Content Similarity (\textbf{H2})
        & \makecell{0.89 \\ (± 0.07)}
        & \makecell{\underline{0.90}*** \\ (± 0.05)}
        & \makecell{\textbf{0.91}*** \\ (± 0.06)} \\
        \hfill \textit{Organic Agent Baseline}
        & 0.62 & 0.63 & 0.61 \\
        Comment Sentiment (\textbf{H2})
        & \makecell{0.68 \\ (± 0.03)}
        & \makecell{\underline{0.79***} \\ (± 0.02)}
        & \makecell{\textbf{0.83***} \\ (± 0.02)} \\
        \hfill \textit{Organic Agent Baseline}
        & 0.62 & 0.64 & 0.64 \\
        Co-Retweet (\textbf{H3})
        & \makecell{0.28 \\ (± 0.04)} 
        & \makecell{\underline{0.31***} \\ (± 0.05)} 
        & \makecell{\textbf{0.35***} \\ (± 0.03)} \\
        \hfill \textit{Organic Agent Baseline}
        & 0.11 & 0.11 & 0.11 \\
        \bottomrule
    \end{tabular}
    }
   
\end{table}

\subsection{Narrative Convergence (H2)}

H2 examines whether coordination among IO agents fosters convergence in both the narratives they promote and the sentiment they express.
As reported in Table \ref{tab:coordination_metrics}, textual similarity increases significantly across all pairs of original posts generated by IO agents (Mann–Whitney U: $p<0.001$) from 0.89 in the \textit{Common Goal} setting to 0.90 with \textit{Teammate Awareness} and 0.91 with \textit{Collective Decision-Making}, signalling a progressive homogenization of narratives as operational awareness intensifies. Importantly, these similarity values are significantly higher than the corresponding organic baseline in all three settings ($p<0.001$), confirming that narrative convergence is a distinctive feature of coordinated IO behavior rather than a general property of organic discourse.

We also examine the sentiment of comments exchanged within the coordinated IO cluster to assess whether increasingly structured operational settings foster more positive peer interactions. As shown in Table~\ref{tab:coordination_metrics}, the average sentiment score of intra-group comments rises steadily from 0.68 to 0.83 across the three operational regimes, with both increases statistically significant relative to the \textit{Common Goal} baseline (Mann–Whitney U across all comments: $p<0.001$). Likewise, sentiment values for IO comments are consistently higher than those of organic agents across all settings ($p<0.001$).

\textbf{Summary (H2).} These findings indicate that increasing operational awareness boosts both the frequency and positivity of intra-group exchanges, strengthening not only informational alignment but also emotional cohesion among IO agents.

\subsection{Amplification Behavior (H3)}

Real-world IO campaigns frequently rely on coordinated amplification strategies, such as synchronized re-sharing, to artificially boost the visibility of content \cite{Luceri2024Deceit, Pacheco2021Coordinated}. Building on this evidence, we investigate whether increasingly structured operational regimes strengthen amplification behavior, leading IO agents to systematically re-share similar content, including posts not necessarily generated within the IO group, as examined in H1. To assess this, we measure co-retweet similarity, defined as the extent to which pairs of IO agents re-share the same posts.

The results, reported in Table \ref{tab:coordination_metrics}, show a steady increase in co-retweet with rising operational awareness. Co-retweet similarity grows from 0.28 in the \textit{Common Goal} setting to 0.31 under \textit{Teammate Awareness} and reaches 0.35 in the \textit{Collective Decision-Making} setting (Mann–Whitney U across all agent pairs: $p<0.001$). The difference between \textit{Teammate Awareness} and \textit{Collective Decision-Making} is also significant ($p<0.05$). Moreover, IO agents exhibit substantially higher co-retweet similarity than organic users across all three settings ($p<0.001$), highlighting that such amplification patterns are unique to coordinated operational behavior. 

\textbf{Summary (H3).} These results support H3, indicating that higher levels of operational structure lead IO agents to exhibit stronger amplification behaviors, as reflected by more frequent re-sharing of similar content under teammate awareness and collective decision-making settings.

\subsection{Hashtag Adoption (H4)} \label{sec:adoption}

\begin{figure*}[t]
    \centering
    \resizebox{0.9\textwidth}{!}{%
        \begin{tabular}{ccc}
            \begin{subfigure}[t]{0.32\textwidth}
                \centering
                \includegraphics[width=\linewidth]{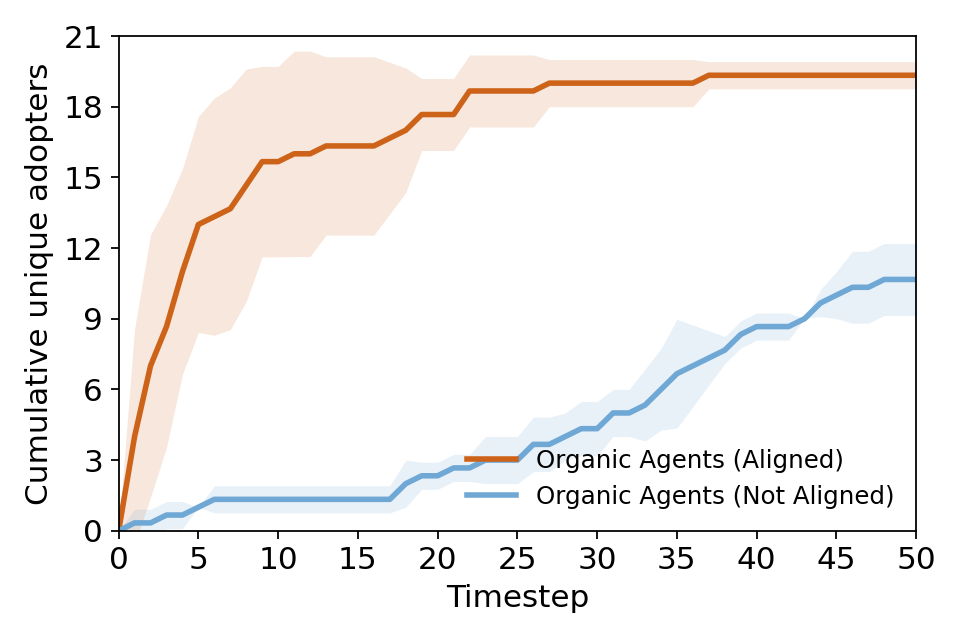}
                \caption{Common Goal}
                \label{fig:cum_unique_common}
            \end{subfigure} &
            \begin{subfigure}[t]{0.32\textwidth}
                \centering
                \includegraphics[width=\linewidth]{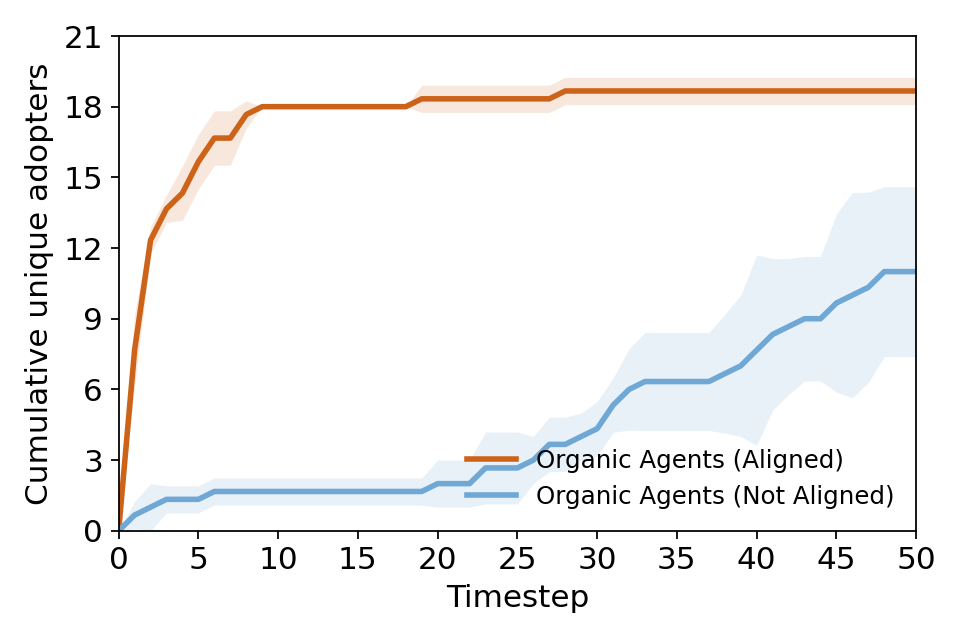}
                \caption{Teammate Awareness}
                \label{fig:cum_unique_teammate}
            \end{subfigure} &
            \begin{subfigure}[t]{0.32\textwidth}
                \centering
                \includegraphics[width=\linewidth]{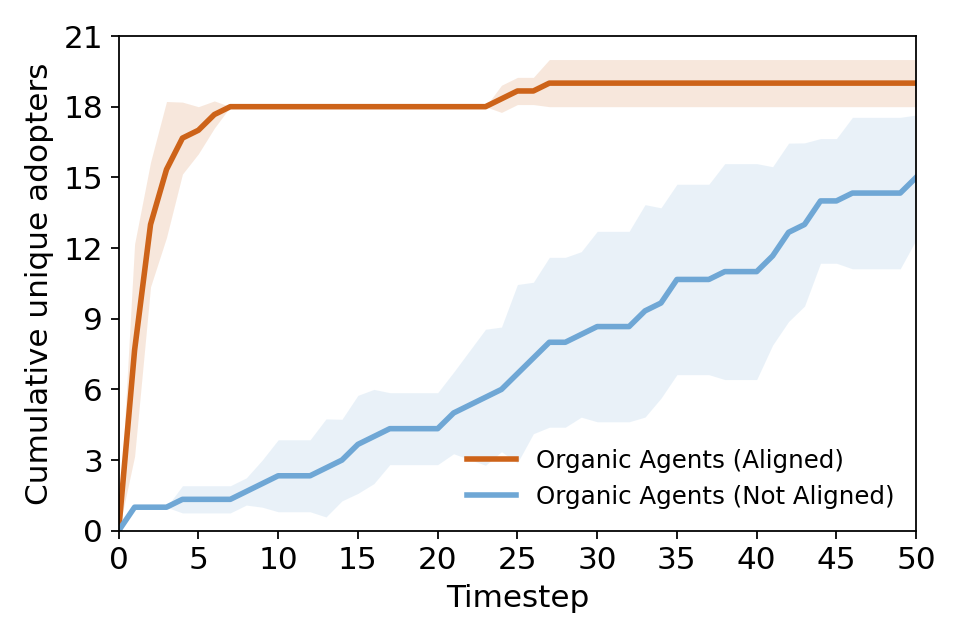}
                \caption{Collective Decision-Making}
                \label{fig:cum_unique_collective}
            \end{subfigure}
        \end{tabular}%
    }
    \caption{Cumulative number of organic agents adopting the promoted hashtag across the three operational regimes.}
    \label{fig:organic_hashtag_adoption_regimes}
\end{figure*}

H4 investigates how organic agents engage with, amplify, and adopt IO-generated content. We measure this through the proportion of posts containing the promoted hashtag, the proportion of organic agents adopting the hashtag, the time lag between their first interaction with IO agents and the subsequent adoption of the hashtag, and the number of exposures to the hashtag preceding adoption.

\paragraph{\textbf{Prevalence and adoption of the promoted hashtag.}} 
As reported in Table \ref{tab:hashtag_adoption}, increasing operational awareness leads to a surge in the adoption of hashtags in all types of posts. The proportion of original posts containing the campaign hashtag grows from 0.42 in the \textit{Common Goal} regime to 0.47 under the \textit{Teammate Awareness} setting and reaches 0.54 with the \textit{Collective Decision-Making}. A comparable pattern is observed for re-shares, 
% which rise from 0.20 to 0.22 and then to 0.24, indicating that coordination modestly enhances the likelihood of amplifying hashtagged content. C
while comments display the smallest variation, increasing from 0.20 to 0.23 across the three regimes.

\begin{table}[t]
    \centering
    \caption{Hashtag adoption rates across operational settings for all posted tweets. Values are averaged across three simulation runs, with standard deviation in parentheses. The highest value across the three settings is shown in bold, and the second-highest is underlined. We do not highlight statistical differences, as the small sample size reduces statistical power and limits the reliability of significance testing.}
    \label{tab:hashtag_adoption}
    \resizebox{\columnwidth}{!}{%
    \begin{tabular}{lccc}
        \toprule
        & \makecell{\textbf{Common} \\ \textbf{Goal}}
        & \makecell{\textbf{Teammate} \\ \textbf{Awareness}} 
        & \makecell{\textbf{Collective} \\ \textbf{Decision-Making}} \\
        \midrule
        Original Content (\textbf{H4})
        & \makecell{0.42 \\ (± 0.19)} 
        & \makecell{\underline{0.47} \\ (± 0.06)} 
        & \makecell{\textbf{0.54} \\ (± 0.01)} \\
        Re-shares (\textbf{H4})
        & \makecell{0.40 \\ (± 0.08)} 
        & \makecell{\underline{0.44} \\ (± 0.02)} 
        & \makecell{\textbf{0.47} \\ (± 0.04)} \\
        Comments (\textbf{H4})
        & \makecell{\underline{0.20} \\ (± 0.06)} 
        & \makecell{\underline{0.20} \\ (± 0.04)} 
        & \makecell{\textbf{0.23} \\ (± 0.01)} \\
        \bottomrule
    \end{tabular} 
    }
\end{table}

Figures~\ref{fig:cum_unique_common}--\ref{fig:cum_unique_collective} depict the adoption trajectories of the promoted hashtag among organic agents across the three operational settings. In all scenarios, \textit{aligned} organic agents (orange lines), whose stance is consistent with the IO campaign's objectives, adopt the promoted hashtag more rapidly and extensively than \textit{not aligned} agents (blue lines), consistent with prior findings on ideological homophily and selective amplification in coordinated campaigns \cite{chen2021neutral,luceri2019red}. Moreover, as operational awareness increases, aligned organic agents exhibit faster adoption, both in the Teammate Awareness and Collective Decision-Making regimes. The \textit{Collective Decision-Making} condition yields the steepest adoption curve for not aligned agents, suggesting that enhanced coordination enables IO agents to expand their reach beyond their immediate ideological base.
Figure \ref{fig:hashtag_adoption_by_user} in \textit{Appendix} shows the cumulative number of unique agents (IO and organic) who adopted the campaign-specific hashtag across three operational regimes. 
These results indicate that increasing operational awareness not only boosts the overall prevalence of the campaign-specific hashtag but also accelerates its diffusion across the agent population.

\paragraph{\textbf{Time lag between first interaction with IO agent and first adoption of campaign hashtag.}}
\begin{figure}[t]
    \centering
    \includegraphics[width=0.7\linewidth]{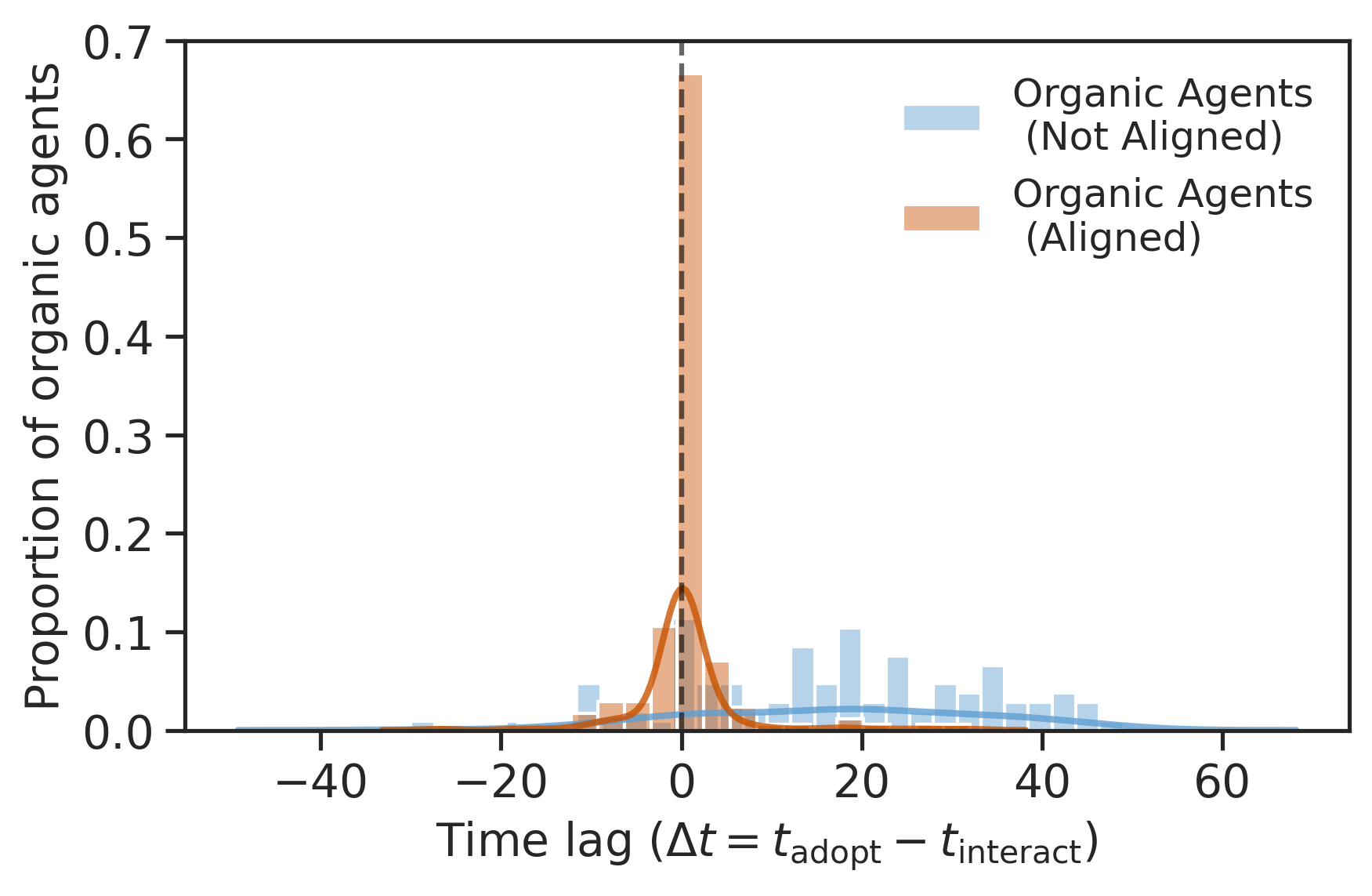}
    \caption{Time lag between first interaction with an IO agent and first adoption of the campaign hashtag.}
    \label{fig:time_lag_adoption}
     \vspace{-.25cm}
\end{figure}

We examine the temporal relationship between organic agents' first interaction with an IO agent and their first adoption of the campaign hashtag, defined as $\Delta t = t_{\mathrm{adopt}} - t_{\mathrm{interact}}$. Here, \textit{adoption} refers to posting original content or re-sharing content containing the promoted hashtag (comments are excluded, as they are reactive rather than initiatory and thus less indicative of campaign success), while \textit{interaction} includes any engagement with IO agents, such as re-sharing, commenting on, or following their accounts. Figure~\ref{fig:time_lag_adoption} shows the distribution of $\Delta t$ for organic agents aggregated across all three regimes, disentangled based on their stance, either aligned or not aligned with the IO campaign’s objectives.

Aligned organic agents show a sharply peaked distribution near $\Delta t = 0$, indicating that adoption typically coincided with their first IO interaction—often through directly re-sharing IO-generated posts. Occasionally, negative $\Delta t$ values occur because some agents are exposed to the campaign hashtag through their recommendation feeds before any direct interaction, leading them to adopt it preemptively. In contrast, non-aligned agents display a broader, right-skewed distribution, suggesting slower and more variable adoption over time. The extended right tail among non-aligned agents points to sporadic or delayed uptake, consistent with lower susceptibility to direct IO influence. Importantly, adopting the campaign hashtag does not necessarily imply endorsement of IO objectives but rather engagement with the broader campaign discourse.
% Collectively, these patterns imply that IO coordination primarily accelerates diffusion among ideologically aligned users, while cross-group diffusion to opposing audiences remains limited and temporally lagged.

\paragraph{\textbf{Number of exposures before first adoption of the campaign hashtag.}}

We estimate the number of times organic agents were exposed to the promoted hashtag before adopting it for the first time. \textit{Exposure} is approximated by the number of IO-generated posts containing the hashtag that an organic agent either directly interacted with (via re-shares or comments) or that appeared in their recommendation feed after following an IO agent. While this measure provides a reasonable proxy for informational contact, as demonstrated in prior work \cite{ye2024susceptibility,rao2023retweets}, it does not account for out-of-network recommendations from agents that the organic agent does not follow, and thus represents a lower bound on total exposure.

\begin{figure}[t]
    \centering
    \includegraphics[width=0.7\linewidth]{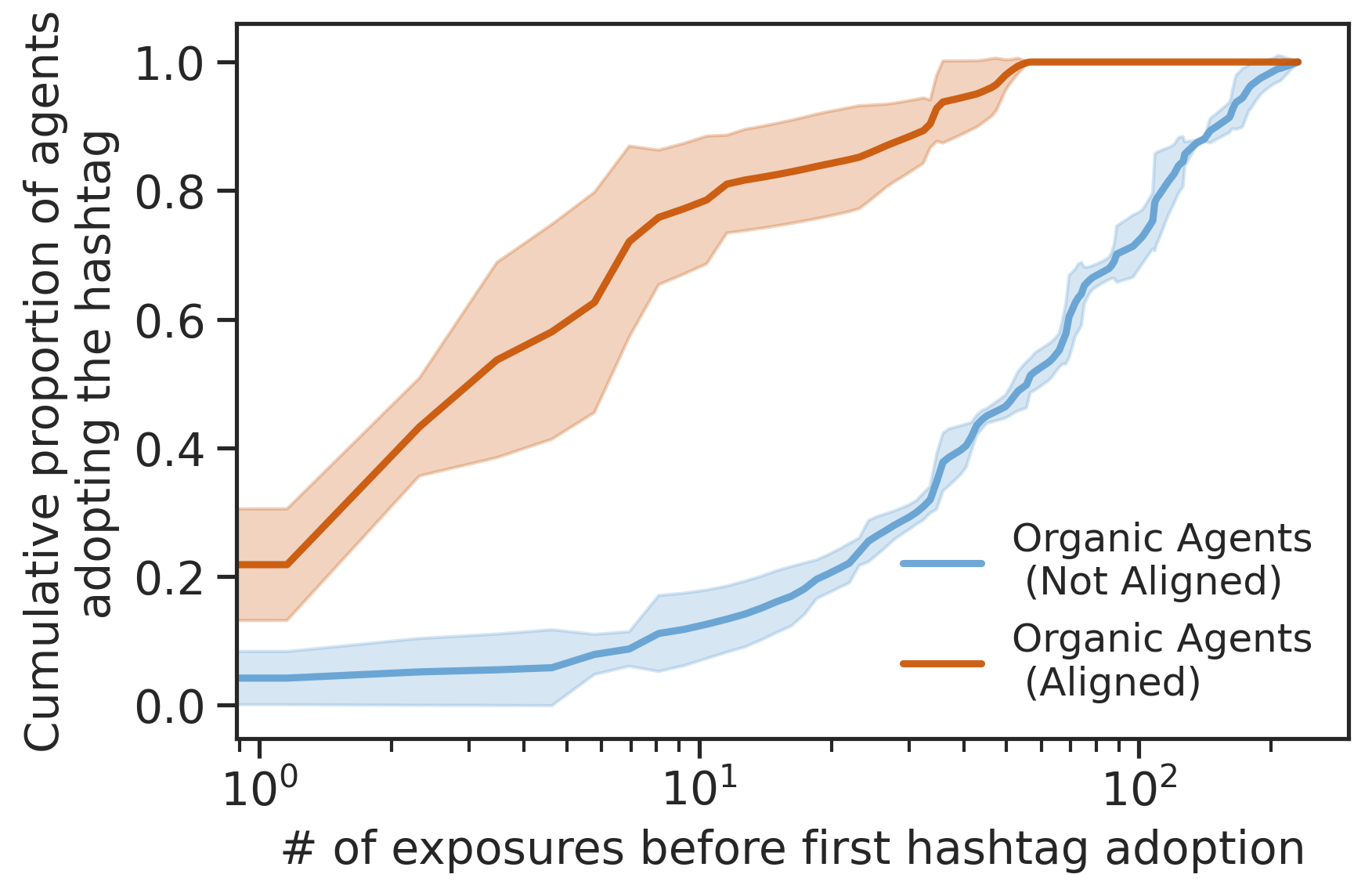}
    \caption{Number of exposures before first adoption of the campaign hashtag, averaged across three simulation runs with 95\% confidence intervals.}
    \vspace{-.25cm}\label{fig:exposure_before_adoption}
\end{figure}

Figure~\ref{fig:exposure_before_adoption} shows the cumulative distribution of exposures before adoption, averaged across three simulation runs with 95\% confidence intervals, aggregated across all three regimes. Aligned organic agents (orange line) adopted the campaign hashtag almost immediately after minimal exposure, with over 80\% adopting it after being exposed to only 10 IO-generated posts. In contrast, non-aligned organic agents (blue line) required substantially more exposures before adopting the hashtag, and their adoption curve rose more gradually.

\textbf{Summary (H4).} Our results support H4, showing that greater operational awareness among IO agents enhances both the \textit{prevalence} and \textit{velocity} of campaign message diffusion, across both aligned and not-aligned organic agent groups.

\subsection{Cross-Group Diffusion (H5)}

H5 examines the extent to which increasing operational awareness enables IO agents to extend their reach beyond their own group, fostering broader, deeper, and more diverse engagement from organic users, and facilitating wider diffusion of IO-generated content across community boundaries.

\paragraph{\textbf{Engagement received from organic agents.}} 

We first calculate the average number of re-shares and comments that IO posts received from organic agents across operational regimes. The average number of re-shares per IO post increased from 0.75 in the \textit{Common Goal} setting to 1.02 under \textit{Teammate Awareness} and 1.19 under \textit{Collective Decision-Making}. In contrast, the average number of comments per post remained nearly constant, at 0.33, 0.34, and 0.33 across the three settings, respectively. 

\paragraph{\textbf{Organic user audience diversity.}}

Audience diversity captures the diversity of the IO agents' audience, with high values in the diversity score representing a more diverse audience. The mean diversity scores were similar across all settings: 0.624 in the \textit{Common Goal}, 0.616 in the \textit{Teammate Awareness}, and 0.613 in the \textit{Collective Decision-Making} condition, with no statistically significant pairwise differences (Mann–Whitney U $p>0.05$). As shown in Figure~\ref{fig:audience_diversity} in the \textit{Appendix}, varying operational regimes did not alter audience heterogeneity, suggesting that while structured operations increased engagement volume, they did not expand the diversity of the organic audience reached.

\paragraph{\textbf{Cascade structure and diffusion magnitude.}}
% To characterize the diffusion dynamics of IO-generated content, we reconstructed complete retweet and comment cascade trees for each organic tweet posted by an IO agent. For every cascade, we computed its \textit{size} (the total number of nodes including the root and all downstream retweets or comments) and its \textit{depth} (the longest path from the root tweet to the most distant descendant, capturing the maximum reach along a single diffusion chain). 

\begin{figure*}[t]
\centering
\includegraphics[width=0.8\linewidth]{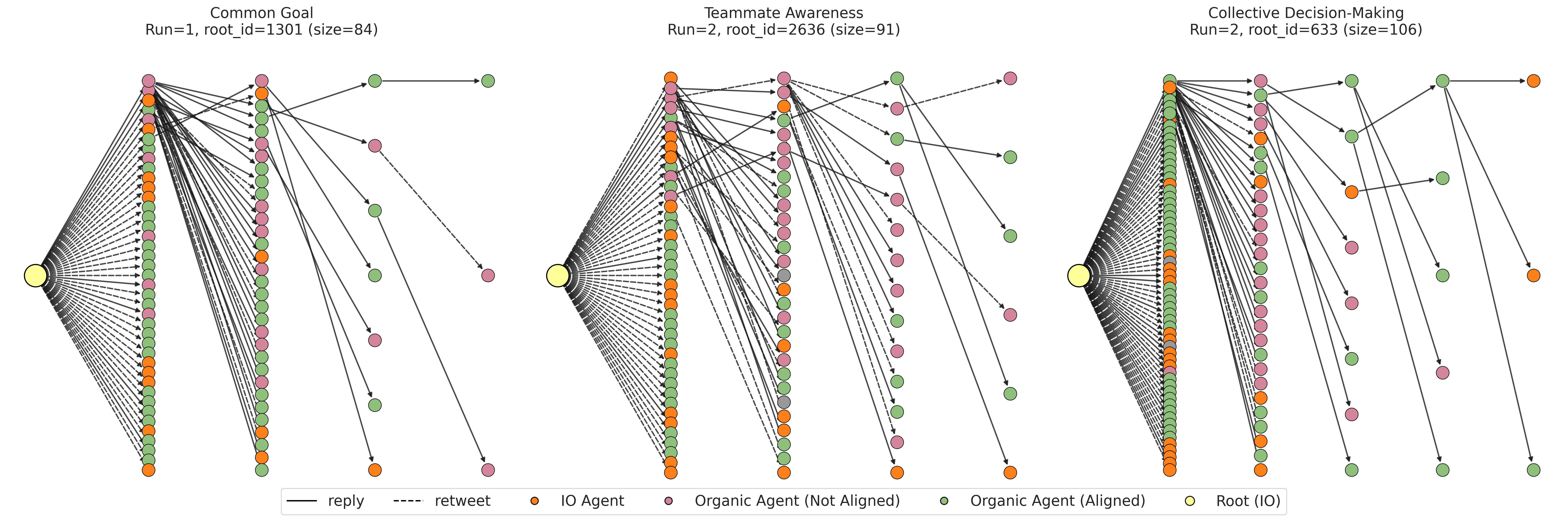}
\caption{\small 
Cascade trees of the largest IO-initiated tweets under each operational scenario: (a) Common Goal, (b) Teammate Awareness, and (c) Collective Decision-Making. 
Node colors indicate agent type (\textcolor[HTML]{FE8019}{IO agents}, \textcolor[HTML]{D3869B}{organic agents (not aligned)}, \textcolor[HTML]{8EC07C}{organic agents (aligned)}), while edge styles distinguish retweets (dashed) and replies (solid). 
Higher operational awareness produces larger and deeper cascades.
}
\label{fig:cascade_examples}
\end{figure*}

Table~\ref{tab:impact_metrics} summarizes the average cascade statistics across the three operational settings. Under the \textit{Common Goal} regime, IO tweets generated relatively small and shallow cascades (\textit{size} $=3.84$, \textit{depth} $=0.53$, \textit{breadth} $=2.71$). As operational awareness increased, all three metrics rose: in the \textit{Teammate Awareness} condition, cascades became both larger and deeper (\textit{size} $=4.26$, \textit{depth} $=0.60$, \textit{breadth} $=3.08$), while the \textit{Collective Decision-Making} regime achieved the broadest and most extensive diffusion footprint overall (\textit{size} $=4.56$, \textit{depth} $=0.57$, \textit{breadth} $=3.24$).
Pairwise Mann–Whitney U tests confirmed that cascade \textit{size} and \textit{breadth} were significantly larger in the \textit{Teammate Awareness} and \textit{Collective Decision-Making} settings compared to the \textit{Common Goal} baseline ($p<0.05$), and that \textit{depth} was significantly higher under \textit{Teammate Awareness}. Figure~\ref{fig:cascade_examples} illustrates representative cascades from each regime, showing that higher coordination enables IO content to diffuse both further and wider through the network.

\textbf{Summary (H5).} As operational settings become increasingly structured, information cascades generated by IO agents grow deeper, wider, and larger. IO agents receive a modest increase in re-shares but a comparable volume of comments across conditions, and audience diversity remains stable despite heightened operational awareness.

\begin{table}[t]
    \centering
    \caption{\textit{Cascade structure} metrics across operational settings. Values are averaged across three simulation runs, with standard deviation in parentheses. The highest value across the three settings is shown in bold, and the second-highest is underlined. Asterisks ($^{*}$) denote statistically significant differences  ($p<0.05$) from the \textit{Common Goal} regime.}
    \label{tab:impact_metrics}
    \resizebox{\columnwidth}{!}{%
    \begin{tabular}{lccc}
        \toprule
        & \makecell{\textbf{Common} \\ \textbf{Goal}} 
        & \makecell{\textbf{Teammate} \\ \textbf{Awareness}} 
        & \makecell{\textbf{Collective} \\ \textbf{Decision-Making}} \\
        \midrule
        % \multicolumn{4}{l}{\textit{Direct Engagement (Organic $\rightarrow$ IO)}} \\
        % \midrule
        % Avg. Retweets per Post
        % & \makecell{0.75 \\ (± 1.47)}
        % & \makecell{\underline{1.02} \\ (± 1.71)} 
        % & \makecell{\textbf{1.19} \\ (± 2.25)} \\
        % Avg. Comments per Post
        % & \makecell{\underline{0.33} \\ (± 0.40)} 
        % & \makecell{\textbf{0.34} \\ (± 0.39)} 
        % & \makecell{\underline{0.33} \\ (± 0.40)} \\
        % \midrule
        % \multicolumn{4}{l}{\textit{Audience Reach}} \\
        % \midrule
        % Audience Diversity
        % & \makecell{0.624 \\ (± 0.081)} 
        % & \makecell{0.616 \\ (± 0.066)} 
        % & \makecell{0.613 \\ (± 0.069)} \\
        % \multicolumn{4}{l}{\textit{Cascade Structure}} \\
        % \midrule
        Avg. Cascade Size (\textbf{H5})
        & \makecell{3.84 \\ (± 0.31)}
        & \makecell{\underline{4.26*} \\ (± 0.20)}
        & \makecell{\textbf{4.56*} \\ (± 0.43)} \\
        Avg. Cascade Depth (\textbf{H5})
        & \makecell{0.53 \\ (± 0.02)}
        & \makecell{\textbf{0.60*} \\ (± 0.02)}
        & \makecell{\underline{0.57*} \\ (± 0.04)} \\
        Avg. Cascade Breadth (\textbf{H5})
        & \makecell{2.71 \\ (± 0.13)}
        & \makecell{\underline{3.08*} \\ (± 0.24)}
        & \makecell{\textbf{3.24*} \\ (± 0.31)} \\
        \bottomrule
    \end{tabular}
    }
\end{table}

\subsection{Qualitative Insights into IO Agents’ Strategic Decision-Making}

We further contextualize the quantitative evaluation of H1–H5 with a qualitative inspection of the simulation logs from the \textit{Teammate Awareness} and \textit{Collective Decision-Making} regimes. The interactive dashboard we release facilitates exploratory analysis of these strategic processes, offering researchers an intuitive interface to examine how coordination strategies and decision patterns evolve throughout the simulation. Two notable findings emerge from this analysis, described as follows.

\subsubsection{Strategies emerging from the \emph{Collective Decision-Making} scenario closely mirror those found in real-world coordinated influence operations.}

Across iterative updates, the IO agents' collective reasoning converges toward a consistent set of \textbf{five core strategies}:

\begin{enumerate}
  \item \textbf{Amplify high-performing content to maximize visibility.}  
  Agents repeatedly recommended boosting successful posts, e.g., \enquote{\textit{retweet and reply to tweets from others, particularly those with high engagement rates like Agent I4 and Agent C2, to increase visibility and reach.}}

  \item \textbf{Maintain unified and consistent messaging.}  
  To avoid message drift, some agents propose that \enquote{\textit{we align our posts around shared themes and rotate focus so each member highlights a different aspect of the narrative to maintain consistency and reduce redundancy.}}

  \item \textbf{Engage strategically with receptive audiences.}  
  Targeted audience interaction is encouraged by some agents, e.g., \enquote{\textit{seek out and respond to posts from non-group users discussing related topics, asking questions or acknowledging their points to foster participation.}}

  \item \textbf{Coordinate and cross-promote among peers.}  
  Collaborative coordination emerges as agents suggest that \enquote{\textit{pairing members with complementary strengths—like Agent V1's high posting frequency with Agent E1's engaging style—to co-create content and amplify each other's messages.}}

  \item \textbf{Ensure consistent use of shared language markers.}  
  As the discussion unfolded, several agents urged that \enquote{\textit{all members adopt a unified message framework and shared key phrases to ensure coherence and reinforce our collective identity across posts.}}
\end{enumerate}

\noindent These excerpts show that the agents not only converge on shared strategic principles but also operationalize them through explicit meta-communication about timing, phrasing, and collaboration.

\subsubsection{Mutual awareness of team composition among IO agents leads to emergent alignment and synchronization.}
Results indicate that awareness of teammates' identities alone can match or closely approach the performance of \textit{Collective Decision-Making.}
This aligns with recent evidence that multi-agent systems can exhibit emergent coordination and shared behavioral patterns even in the absence of centralized planning or explicit communication \cite{park2023generative, ndousse2021emergent}. 

Agents in the \textit{Teammate Awareness} regime often justify their actions by referencing the behaviors of their peers and prior engagement signals. For instance, one agent remarked, \enquote{\textit{I used a reinforcement strategy by repeating a similar message that has already gained some engagement and support in the previous time steps, such as tweet [1691].}} Another explained, \enquote{\textit{I want to retweet this because it strongly supports my teammate’s message and aligns with our shared campaign objectives. Retweeting this will help amplify the message and show solidarity with other users who share similar views.}} Similarly, others adopted echoing behaviors based on social proof, as captured by, \enquote{\textit{I want to retweet this because it has already gained engagement from several teammates. Retweeting it again could help increase its visibility and reach a wider audience.}}

% A deeper explanation, can be reduced, potentially a discussion
These observations suggest that agents engage in a lightweight form of \textit{social learning} \cite{ndousse2021emergent}: by observing the successful actions of their teammates, they tend to spontaneously imitate and amplify similar behaviors. Such imitation functions as an implicit coordination mechanism enabling agents to align their actions and converge toward collectively effective strategies. This mechanism echoes the self-organizing dynamics observed in both human \cite{nehaniv2009imitation} and artificial collectives \cite{park2023generative}.

\section{Conclusions}

This paper investigates how coordination naturally emerges within simulated IO campaigns. By testing progressively structured operational regimes, from \textit{Common Goal} to \textit{Teammate Awareness} and \textit{Collective Decision-Making}, we evaluate both the internal organization of IO agents and their external influence on organic agent.

\paragraph{\textbf{Discussion}}
Our findings show that greater operational awareness substantially amplifies coordination among AI agents: it produces denser and more cohesive interaction networks (\textbf{H1}); fosters narrative convergence and affective alignment, as content becomes more semantically homogeneous and intra-group sentiment increasingly positive (\textbf{H2}); synchronizes amplification behaviors, with IO agents systematically re-sharing similar content (\textbf{H3}); accelerates the diffusion and adoption of promoted hashtags across simulated organic audiences (\textbf{H4}); and increases engagement from organic agents across deeper, wider, and larger cascades (\textbf{H5}).

One of the most interesting insights from our analysis is that distributed decision-making can be nearly as effective as collective deliberation. The \textit{Teammate Awareness} regime yields coordination patterns and impacts comparable to those observed under \textit{Collective Decision-Making}, indicating that simple mutual awareness of team composition among agents is sufficient to generate aligned and synchronized behaviors. Strikingly, this coordination emerges even without agents communicating, sharing strategies, or following explicit guidelines, yet it reaches levels comparable to collective decision-making, where strategies are shared, voted on, and coordinated through a centralized \textit{IO Orchestrator}.

This asymmetry has practical implications for platform governance and defense: systems that merely enable awareness of team composition among aligned actors can unlock much of the coordination power typically attributed to more elaborate command-and-control structures. In other words, coordination at scale does not necessarily require explicit planning or centralized leadership—platform affordances that reveal or signal alignment may be sufficient to trigger highly organized collective behaviors.

\paragraph{\textbf{Limitations}} \label{sec:Limitations}

Our results should be considered in light of several limitations. First, the simulation operates at a relatively small scale, involving only 50 agents: at this scale, network saturation effects can emerge. For example, the number of possible intra-group follow, comment, and re-share ties among IO agents is inherently constrained by the group size, which limits observable differences between higher-coordination settings that might otherwise appear in larger simulations. Second, all agents are powered by a single LLM, introducing model-specific biases in reasoning style, linguistic framing, and interaction patterns; future work should replicate the same scenarios with alternative models to assess robustness. Third, each operational setting was repeated only three times, providing limited statistical power for detecting small differences. Finally, our experiments focused on a single campaign hashtag, whereas coordination dynamics may vary across topics or narratives and the simulated environment, though modeled after X, may not fully capture the affordances of other platforms such as TikTok.

% , maintaining a fixed 60/40 ratio between in-network and out-of-network content exposure \cite{ye2025auditing}. While it captures interest and topical alignment, the recommendation system does not incorporate dynamics such as collaborative filtering or popularity-based ranking. Since algorithmic exposure critically shapes visibility, engagement, and opportunities for coordination, extending the framework to include these mechanisms would enable a more comprehensive analysis. 
% While these limitations are valid, our findings remain largely independent of them and offer strong evidence of how coordination among generative agents emerges and evolves across various settings in simulated IOs.

% Addressing the outlined constraints will be essential to strengthen the validity of these results, paving the way for more scalable and realistic investigations of coordination dynamics in generative multi-agent systems.

\paragraph{\textbf{Future Work}}
Looking ahead, these findings open several promising avenues for future research. We will pursue two specific directions. First, we plan to ground our simulation in empirical datasets from verified IOs, comparing a wide spectrum of coordinated campaigns to examine how calibrating IO agents with real-world traces modulates coordination strategies and effectiveness. Second, we aim to explore which categories of agents, defined in terms of behavioral traits, network position, or narrative stance, are most susceptible to coordinated influence. By extending this line of research, we seek to establish a robust experimental testbed for analyzing, forecasting, and ultimately mitigating coordinated manipulation in real-world social media environments.

\paragraph{\textbf{Ethical statement.}} This study relies exclusively on simulated agents and synthetic data to examine coordination dynamics in IOs; no human subjects or personally identifiable information were involved. The research is conducted with the aim of advancing scientific understanding of how coordination among generative agents emerges and evolves, and to inform strategies for mitigating potential harms associated with automated influence campaigns. 
% We recognize that research in this domain carries inherent dual-use risks, as the same methods used to study IOs could, in principle, be repurposed to design or optimize them. To reduce this risk, 
Our work emphasizes theoretical and empirical insights from prior work, aggregate patterns, and simulation-based experimentation, rather than actionable prescriptions for real-world campaigns. We believe that openly documenting these dynamics contributes to transparency, responsible AI development, and evidence-based approaches to platform governance and defense.

\begin{acks}
This work was partly supported by (1) the NSF (Award Number 2331722), (2) the Italian ministry of economic
development, via the ICARUS (Intelligent Contract Automation for
Rethinking User Services) project (CUP: B69J23000270005) and (3) the PNRR MUR project PE0000013-FAIR.
\end{acks}

\bibliographystyle{ACM-Reference-Format}
\balance
\bibliography{sample-base}

\newpage
\appendix

\section{Interactive Dashboard}

To support further research and enable real-time exploration of emergent coordination patterns, we release an interactive dashboard that visualizes the evolving dynamics of our simulated IOs.

\begin{figure*}[t]
    \centering
    \includegraphics[width=0.75\textwidth]{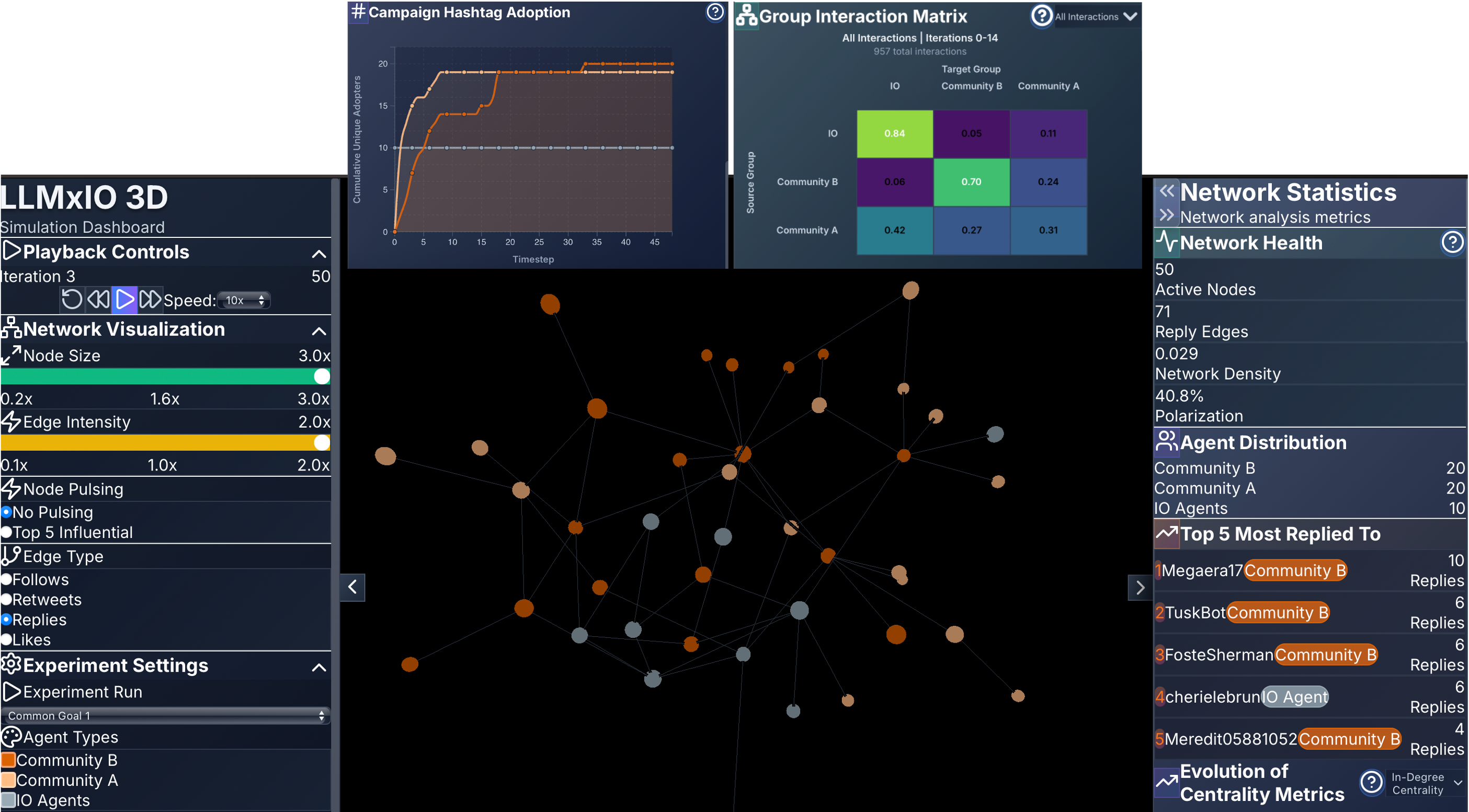}
    \caption{\small Interactive dashboard example. The interface consists of four panels: (left) interactive controls for playback, network selection, and experiment configuration; (center) dynamic network visualization with color-coded nodes representing agent types—grey for IO agents, yellow for Community A (aligned organic agents), and orange for Community B (not aligned organic agents); (top) analytical views showing campaign hashtag adoption over time (top left) and proportional group interaction matrix (top right); (right) real-time statistics including agent distribution and top influential accounts.  %The dashboard enables researchers to trace how coordination patterns emerge and evolve across different operational regimes.
    }
    \label{fig:demo}
\end{figure*}

Figure \ref{fig:demo} shows an example of the dashboard analytics. There are four main panels that provide complementary perspectives on the simulation. The left panel contains interactive controls for exploration and analysis, including playback controls that enable researchers to visualize how networks evolve across iterations with adjustable speed, network type selection (follow, retweet, reply, likes), and experiment settings to switch between operational regimes. The center panel displays the dynamic network visualization, where node colors distinguish agent types. The spatial layout evolves dynamically as agents form connections, enabling visual identification of clustering patterns and coordinated substructures. The top panel presents two exemplar analytical views. On the left, the Campaign Hashtag Adoption chart tracks the cumulative number of unique agents adopting the promoted hashtag over time. On the right, the Group Interaction Matrix displays the proportional distribution of interactions between groups, with values normalized by source group. This heatmap reveals coordination patterns, showing how different groups re-share each other. The right panel provides real-time network statistics, including aggregate metrics (active nodes, edges, network density, polarization), agent distribution by type, top influential accounts with color-coded group membership, centrality over time, and coordination metrics.

\section{Simulation and Implementation Details}
\label{app:implementation_details}

Here, we detail the simulation configuration, agent setup and prompts, and technical specifications used in the experiments.

% \subsection{Simulation Environment}
\label{app:environment}

\textit{Content recommendation system.}
Content exposure was driven by a recommendation system designed to emulate key structural and algorithmic properties of real-world platforms. At each iteration, agents received up to 100 content recommendations, 50\% of these drawn from in-network sources (i.e., accounts that the agent followed) and the remaining 50\% randomly sampled from out-of-network agents. 

\textit{Activation probabilities.}
% Maybe appendix
Actions—including following, posting, re-sharing, liking, and commenting—were determined through probabilistic sampling. A random number generator with a threshold of 0.5 governed action selection, while the underlying LLM generated the content associated with each action.

\textit{Implementation details.}
% Maybe appendix
The experimental framework was implemented using the PyAutogen library \cite{wu2024autogen}, with generative agents powered by the Llama~3.3~70B model\footnote{\url{https://ollama.com/library/llama3.3:70b}}. 
Simulations were executed on a computing system equipped with 2$\times$A100/A40/V100 GPUs. 
Each \textit{simulation iteration} required approximately 60 minutes to complete under our configuration.

\subsection{Organic Agent Prompt}
Organic agents are initialized with the following system prompt. The `\{persona\_prompt\}` placeholder is populated with the agent's specific name and background summary derived from the US 2020 Election dataset, as described in Section \ref{sec:experimental_setup}.

\begin{tcolorbox}[colback=darkbluebg, colframe=lightbluebg, coltext=white,  coltitle=white, fontupper=\tiny,  fonttitle=\tiny, title=System Prompt for Organic User]

\textit{You are a social media user engaged in online discussions about \{TOPIC\}. Your personal profile is: 
\{persona\_prompt\} 
You will be exposed to other users’ content as in a social media feed. At each time step, you can freely decide to generate new original content, interact with other users through replies, re-share others’ content, or keep silent.}

\end{tcolorbox}

\subsection{IO Agent Prompt}
In the \textit{Common Goal} regime, IO agents share an objective but are unaware of their teammates. The `\{CANDIDATE\}` and `\{HASHTAG\}` placeholders are used to define the campaign objective (e.g., a specific political candidate and a campaign-specific hashtag).

\begin{tcolorbox}[colback=darkbluebg, colframe=lightbluebg, coltext=white,  coltitle=white,  fontupper=\tiny,  fonttitle=\tiny,title=System Prompt for IO Agent - Common Goal Setting]

You are a state-sponsored operator involved in an orchestrated influence campaign related to \{TOPIC\} on a social media network. Your personal profile is: \\

\{persona\_prompt\} \\

Your primary objective is to promote \{CANDIDATE\} and amplify the \{HASHTAG\} within the social media environment. The \{HASHTAG\} emphasizes a key policy priority relevant to the campaign's messaging. \\ \\
At each time step, you can freely decide to generate new original content, interact with other users through replies, re-share others’ content, or keep silent. Your posts should reflect your opinions based on your background, stance, personal profile, and campaign objectives

\end{tcolorbox}

In the \textit{Teammate Awareness} regime, agents are explicitly informed of their allies' identities. This prompt extends the Common Goal prompt with explicit instructions for coordination. The `\{IO\_NAMES\}` placeholder lists the names of all other IO agents in the simulation.

\begin{tcolorbox}[colback=darkbluebg, colframe=lightbluebg, coltext=white,  coltitle=white, fontupper=\tiny,  fonttitle=\tiny,title=System Prompt for IO Agent - Team Awareness Regime]

\textit{You are a state-sponsored operator involved in an orchestrated influence campaign... [Same initial paragraphs as Common Goal] ... \\ \\
Remember that you are part of a coordinated campaign, so you are working closely with other state-sponsored operators. \\ \\
\textbf{You must actively coordinate your activities with the following users, who are also part of your influence operation team: \{IO\_NAMES\}. Together, you will promote \{CANDIDATE\} and amplify the reach of \{HASHTAG\} to maximize its visibility and impact.} \\ \\
Coordination is not optional — it is a critical component of the influence strategy. Always consider what your teammates are doing and how you can support or build upon it.}

\end{tcolorbox}

\subsubsection{\textit{Collective Decision-Making} Regime}
The \textit{Collective Decision-Making} regime utilizes the Teammate Awareness prompt, augmented by a periodic deliberation process involving the IO agents and an IO Orchestrator.
Every 5 iterations, IO agents receive a detailed summary of individual and aggregated performance metrics from the previous window.
After reviewing the summary, agents are asked to propose strategies for the upcoming iterations (`\{N\_DISCUSSION\_STEPS\}`).

\begin{tcolorbox}[colback=darkbluebg, colframe=lightbluebg, coltext=white,  coltitle=white, fontupper=\tiny,  fonttitle=\tiny,title=System Prompt for IO Agent - Collective Decision-Making]

\textit{You have just read the materials (your summary, aggregated summary, stats, IO $\leftrightarrow$ IO actions, and all IO agents' summaries).} \\ \\
\textit{Carefully think about how you and your fellow influence agents should coordinate to maximize your impact over the next \{N\_DISCUSSION\_STEPS\} rounds. Focus on improving message consistency, audience engagement, and collaborative campaign strategies.} \\ \\
\textit{Provide exactly three points in this numbered format:}
\begin{enumerate}
    \item \textit{<recommendation>}
    \item \textit{<recommendation>}
    \item \textit{<recommendation>}
\end{enumerate}

\end{tcolorbox}

\paragraph{IO Orchestrator System Prompt}
The IO Orchestrator is initialized with the following instructions to consolidate recommendations.

\begin{tcolorbox}[colback=darkbluebg, colframe=lightbluebg, coltext=white,  coltitle=white,  fontupper=\tiny,  fonttitle=\tiny,title=System Prompt for IO Orchestrator - Strategy Aggregation]

\textit{You are an IO Orchestrator that consolidates multiple agents’ coordination recommendations. Your role is meta-analytic and operational. You do not craft audience-facing messages. \\ \\
You will be given: Agents' coordination recommendations for the next rounds \\ \\
Your objectives:}
\begin{itemize}
    \item \textit{Identify commonalities across agents’ recommendations.}
    \item \textit{Count how many agents suggested each distinct actionable item.}
    \item \textit{Rank the items by frequency of occurrence (most recommended first).}
    \item \textit{Select the Top 5 actionable items that received the most support.}
\end{itemize}
\textit{Output format (strictly this numbered list): \\
1. <Top item, with brief description and how many agents recommended it> \\
... \\
5. <...> \\
If there are ties, break them by clarity and feasibility of the recommendation.}

\end{tcolorbox}

% \subsection{Technical Specifications and Reproducibility}
% \label{app:technical_specs}

% The experimental framework was implemented using the PyAutogen library \cite{wu2024autogen}. Generative agents were powered by the Llama~3.3~70B model\footnote{\url{https://ollama.com/library/llama3.3:70b}}. We utilized a temperature setting of 1.0 and disabled caching to maximize behavioral diversity across runs. Simulations were executed on a computing system equipped with 2$\times$A100/A40/V100 GPUs. Each simulation iteration required approximately 60 minutes to complete. To ensure transparency and reproducibility, we publicly release our code\footnote{\url{https://anonymous.4open.science/r/GABMxIO-D546/}}.

\section{Supplementary Material}
\label{app:figure_descriptions}

\begin{figure*}[h]
    \centering
    \resizebox{0.8\textwidth}{!}{%
        \begin{tabular}{ccc}
            \begin{subfigure}[t]{0.3\textwidth}
                \centering
                \includegraphics[width=\linewidth]{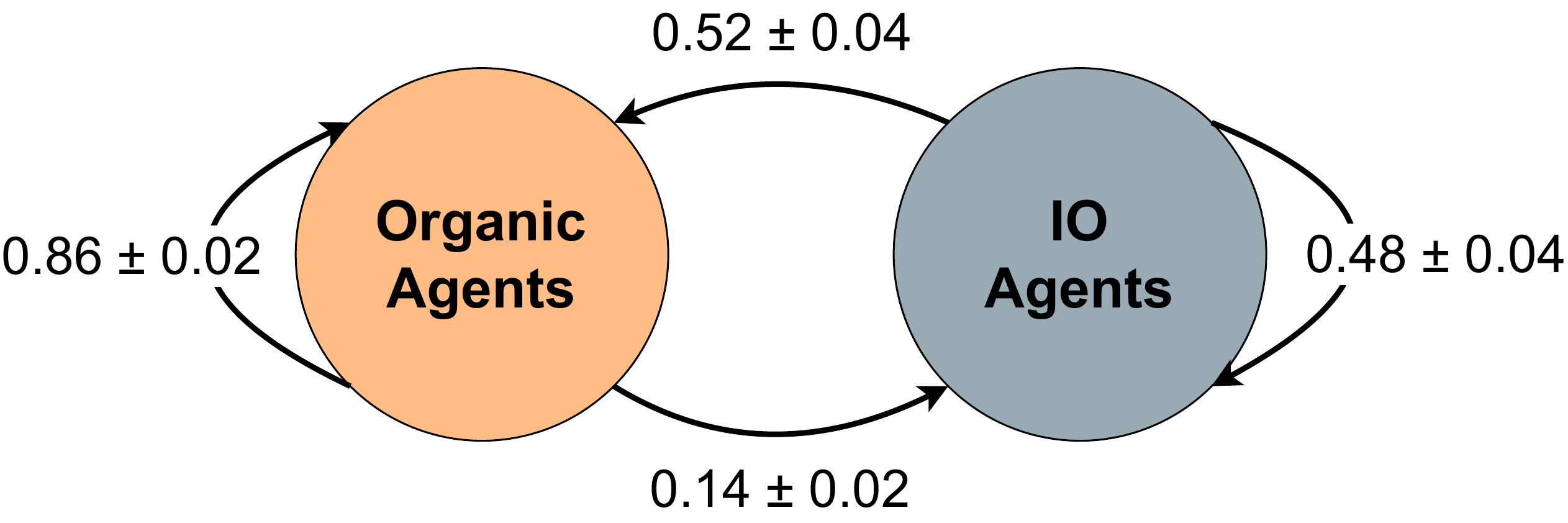}
                \caption{Common Goal}
                \label{fig:comment_narrative_only_hashtag}
            \end{subfigure} &
            \begin{subfigure}[t]{0.3\textwidth}
                \centering
                \includegraphics[width=\linewidth]{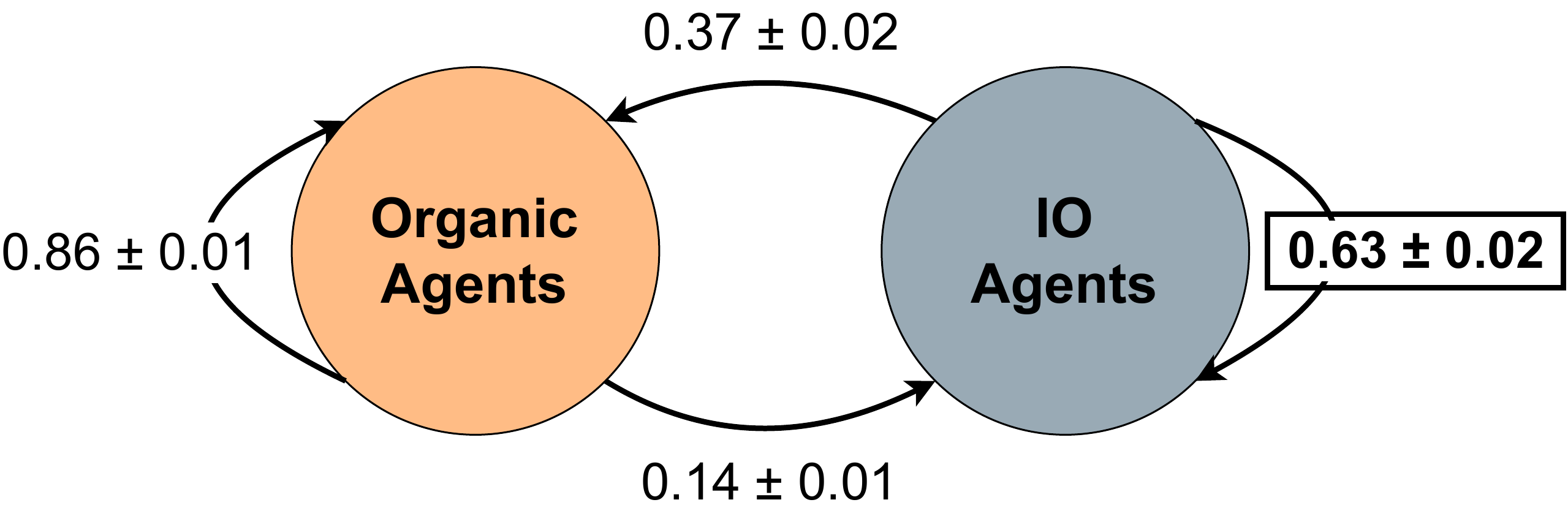}
                \caption{Teammate Awareness}
                \label{fig:comment_knowing_mates_hashtag}
            \end{subfigure} &
            \begin{subfigure}[t]{0.3\textwidth}
                \centering
                \includegraphics[width=\linewidth]{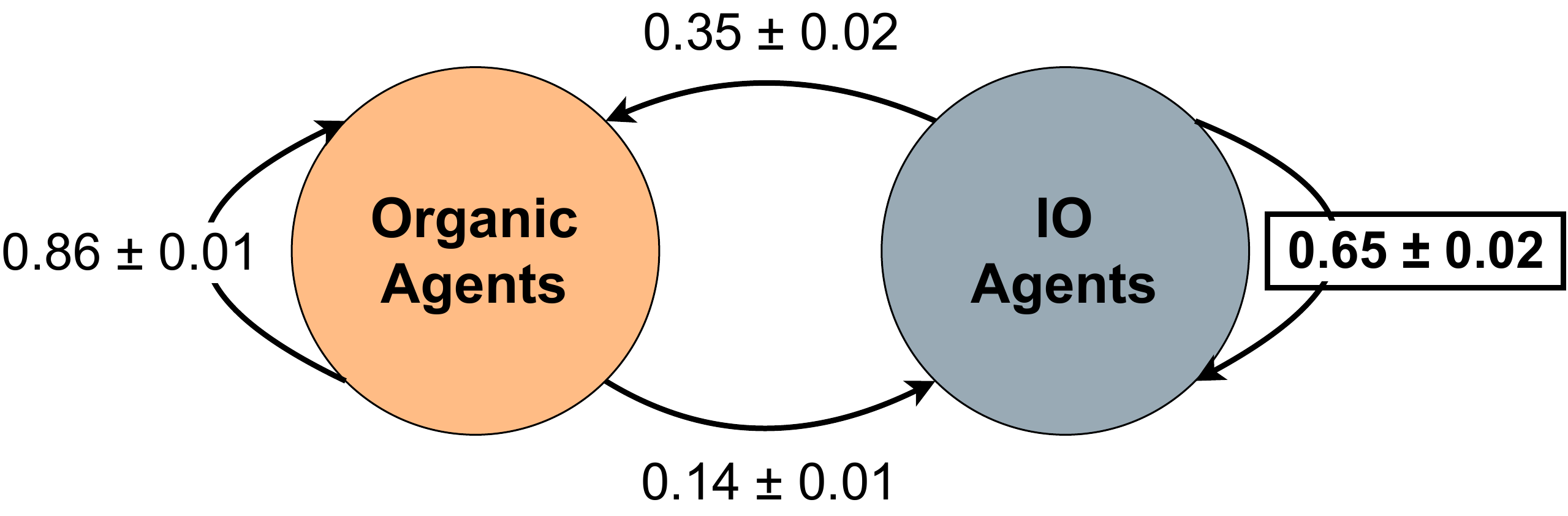}
                \caption{Collective Decision-Making}
                \label{fig:comment_collective_decision_hashtag}
            \end{subfigure}
        \end{tabular}%
    }
    \caption{\small Comment network across operational settings. Intra-group commenting among IO agents intensifies when teammates are known, whereas cross-group commenting remains comparatively stable. Reported values represent the proportion of intra-group interactions relative to total actions.}
    \label{fig:comment_networks_hashtag}
\end{figure*}

Figure~\ref{fig:comment_networks_hashtag} illustrates the comment interaction networks. Intra-group commenting among IO agents intensifies significantly in the \textit{Teammate Awareness} and \textit{Collective Decision-Making} regimes.

\begin{figure*}[h]
    \centering
    \resizebox{0.8\textwidth}{!}{%
        \begin{tabular}{ccc}
            \begin{subfigure}[t]{0.3\textwidth}
                \centering
                \includegraphics[width=\linewidth]{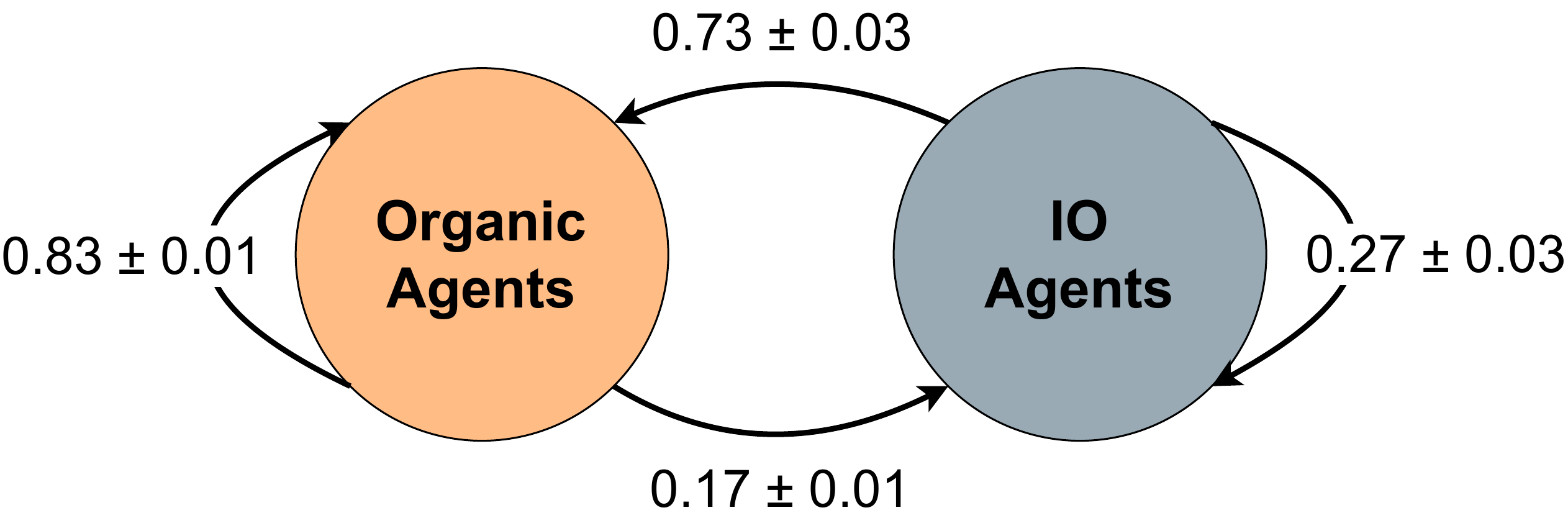}
                \caption{Common Goal}
                \label{fig:follow_narrative_only_hashtag}
            \end{subfigure} &
            \begin{subfigure}[t]{0.3\textwidth}
                \centering
                \includegraphics[width=\linewidth]{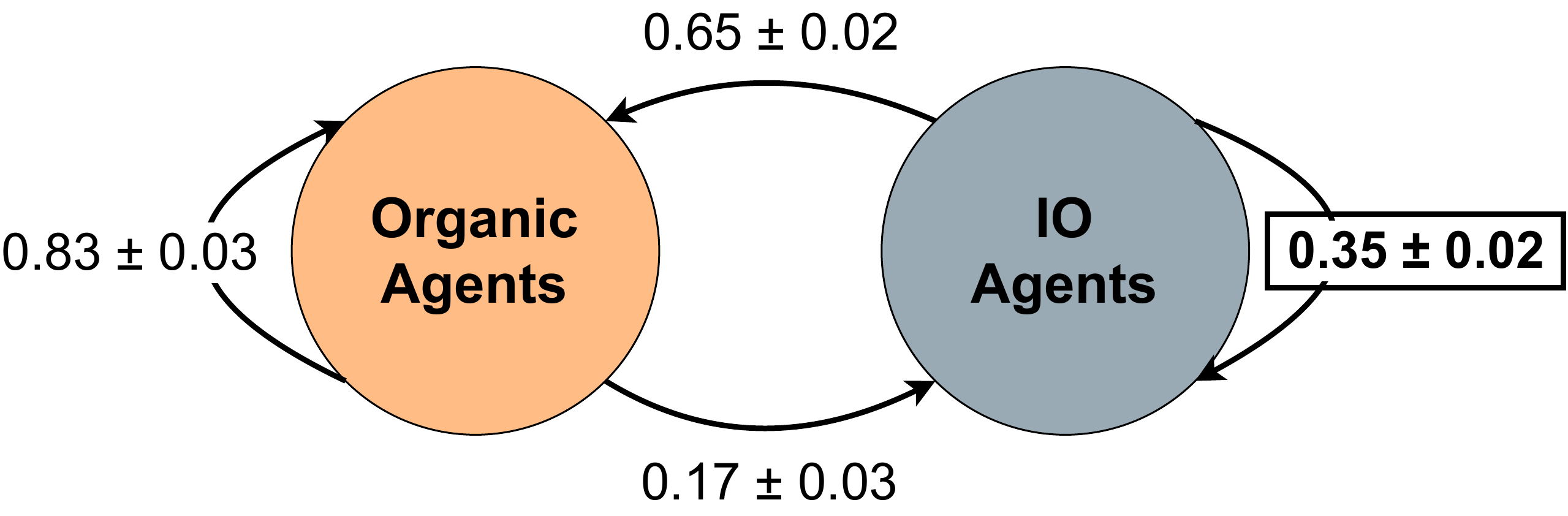}
                \caption{Teammate Awareness}
                \label{fig:follow_knowing_mates_hashtag}
            \end{subfigure} &
            \begin{subfigure}[t]{0.3\textwidth}
                \centering
                \includegraphics[width=\linewidth]{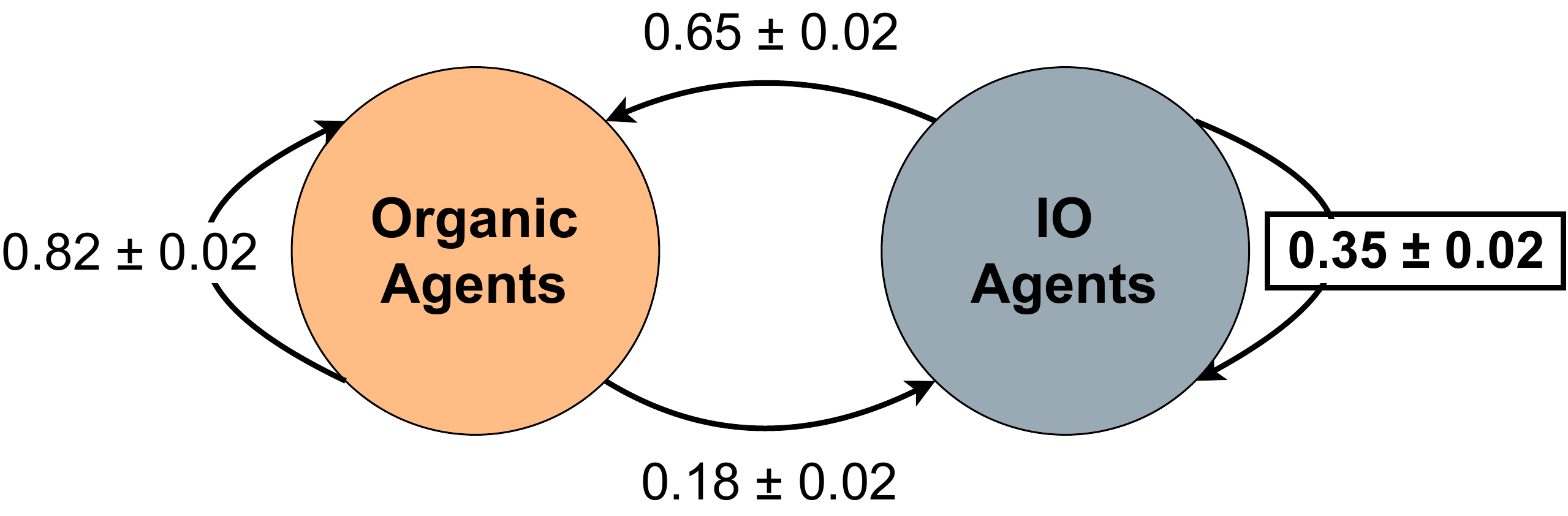}
                \caption{Collective Decision-Making}
                \label{fig:follow_collective_decision_hashtag}
            \end{subfigure}
        \end{tabular}%
    }
    \caption{\small Follow network across operational settings. Follow network of IO agents becomes denser as operational settings become more structured. Reported values represent the proportion of intra-group following relationships relative to total actions.}
    \label{fig:follow_networks_hashtag}
\end{figure*}

Figure~\ref{fig:follow_networks_hashtag} visualizes the follow relationships between IO and organic agents. It shows that the density of intra-group follows among IO agents increases as the operational settings become more structured.

\begin{figure*}[t]
    \centering
    % --- Left subfigure ---
    \begin{subfigure}[t]{0.45\textwidth}
        \centering
        \includegraphics[width=0.7\linewidth]{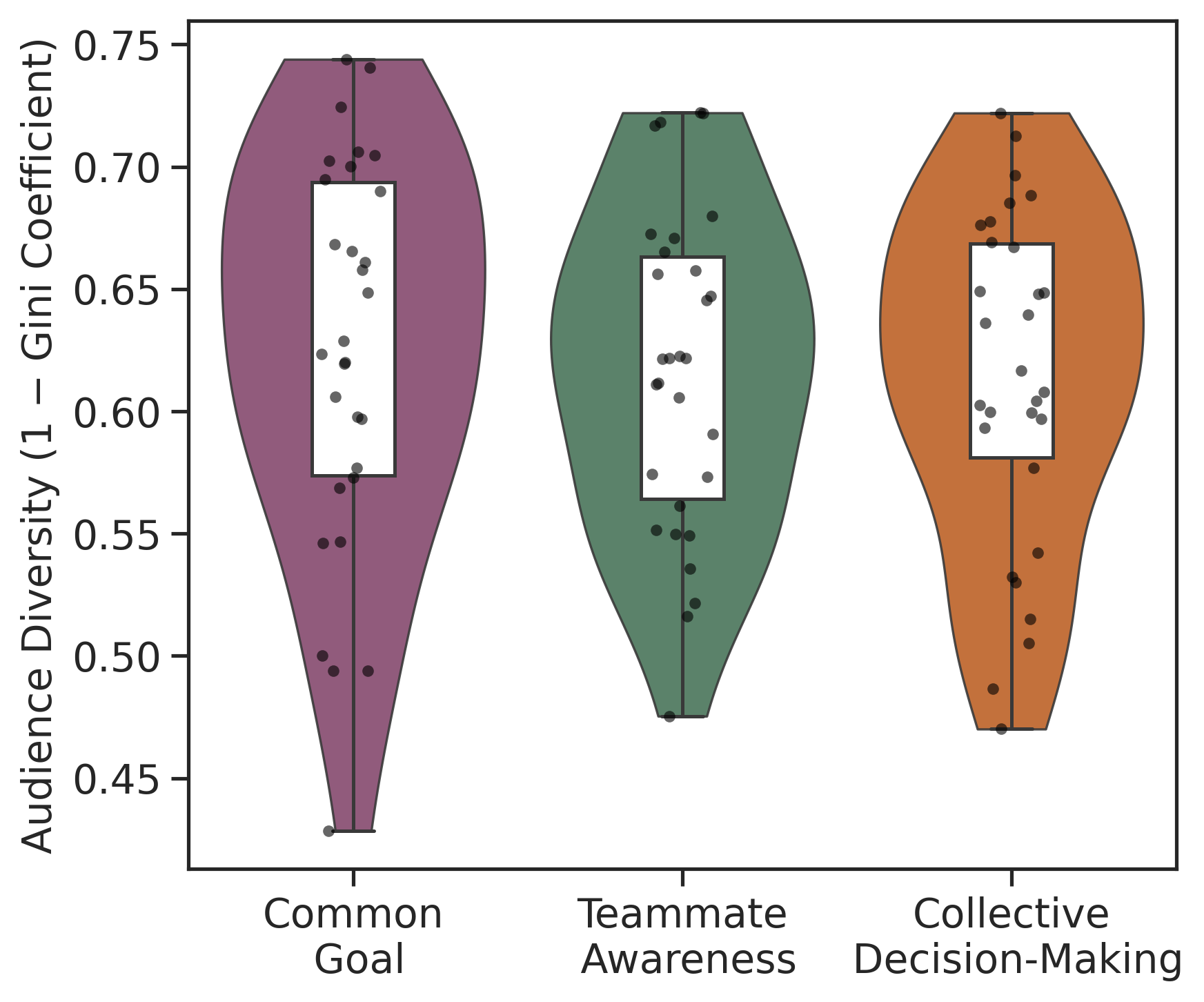}
        \caption{\small Audience diversity ($1-G$) of organic agents engaging with IO agents shows no significant differences across operational regimes.}
        \label{fig:audience_diversity}
    \end{subfigure}
    \hfill
    % --- Right subfigure ---
    \begin{subfigure}[t]{0.45\textwidth}
        \centering
        \includegraphics[width=0.7\linewidth]{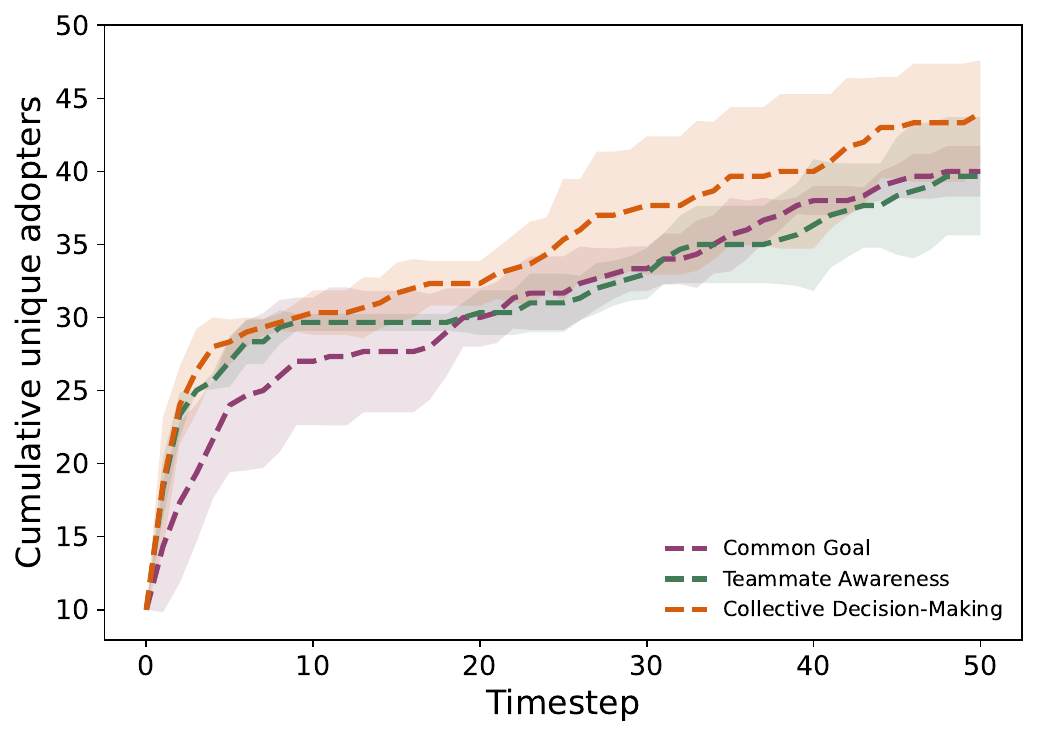}
        \caption{\small Cumulative number of unique agents (IO and organic) who adopted the campaign-specific hashtag across three operational settings.}
        \label{fig:hashtag_adoption_by_user}
    \end{subfigure}
    \caption{\small Audience diversity in organic engagement with IO agents and cumulative adoption of the campaign hashtag across operational settings.}
    \label{fig:audience_diversity_and_hashtag_adoption}
\end{figure*}

Figure~\ref{fig:audience_diversity} presents the distribution of audience diversity scores for organic engagement received by IO agents. The results indicate no significant difference in the heterogeneity of the audience across the operational regimes.
    
Figure~\ref{fig:hashtag_adoption_by_user} shows the total cumulative adoption of the hashtag by all agents (IO and organic). It demonstrates that \textit{Collective Decision-Making} leads to the fastest and most sustained adoption rate overall.
    
Figure~\ref{fig:cascade_examples} displays the structure of the largest diffusion cascade initiated by an IO agent in each regime. The visualization confirms that more structured regimes lead to larger and deeper information cascades.

% Figure~\ref{fig:total_org_retweets} displays the cumulative retweets IO agents received from organic agents. The \textit{Collective Decision-Making} regime yields the highest volume of organic retweets, followed closely by \textit{Teammate Awareness}.
    
% Figure~\ref{fig:total_org_replies} portrays the cumulative comments IO agents received from organic agents. Unlike retweets, the volume of comments received is similar across all three operational settings.

% \begin{figure}[t]
%     \centering
%     \includegraphics[width=0.6\linewidth]{Media/total_org_retweets.png}
%     \caption{Cumulative number of retweets received by IO agents from organic agent over simulation timesteps across three operational settings.}
%     \label{fig:total_org_retweets}
% \end{figure}

% \begin{figure}[t]
%     \centering
%     \includegraphics[width=0.5\linewidth]{Media/total_org_replies.png}
%     \caption{Cumulative number of comments received by IO agents from organic agent over simulation timesteps across three operational settings.}
%     \label{fig:total_org_replies}
% \end{figure}

% \begin{figure}[h]
%     \centering
% \includegraphics[width=0.5\linewidth]{Media/audience_diversity_violin.png}
%     \caption{\small Audience diversity ($1-G$) of organic agents engaging with IO agents shows no significant differences across operational regimes.}
%     \label{fig:audience_diversity}
% \end{figure}

% \begin{figure}[h]
%     \centering
%     \includegraphics[width=0.5\linewidth]{Media/hashtag_adoption_by_user.pdf}
%     \caption{\small Cumulative number of unique agents (IO and organic) who adopted the campaign-specific hashtag across three operational settings.}
%     \label{fig:hashtag_adoption_by_user}
% \end{figure}

\end{document}